\pdfoutput=1 
\documentclass[aps,prc,
twocolumn,eqsecnum,nofootinbib
]{revtex4-1}

\usepackage[]{qrcode}
\usepackage{graphicx,subfigure,latexsym}
\usepackage{amsmath,amssymb,amsfonts}
\usepackage{bm}
\usepackage{hyperref}
\usepackage{ulem}
\usepackage{color}
\usepackage{lineno}

\usepackage{soul}
\setulcolor{blue}
\setstcolor{red}
\sethlcolor{yellow}

\def \bes {\begin{subequations} }
\def \ees {\end{subequations}}

\def \be {\begin{equation}}
\def \ee{\end{equation}}

\def \pd {\partial}

\def \eq {Eq.~}

\def \b {\beta}

\def \d {\delta}
\def \e {\epsilon}

\def \k {\kappa}

\def \l {\lambda}

\def \s {\sigma}
\def \t {\tau}
\def \th {\theta}
\def \x {\xi}

\def \O {\Omega}

\def \<{\langle}
\def \>{\rangle}
\def \+{\dagger}
\def \({\left(}
\def \){\right)}
\def \[{\left[}
\def \]{\right]}

\def \vx {\bm{x}}

\def \tk {\tilde{\kappa}}
\def \th {\tilde{h}}
\def \tM {\tilde{M}}
\def \tl {\tilde{\lambda}}

\def \tM{\tilde{M}}

\def \tsigma{\tilde{\sigma}}

\def \equ {\text{eq}}

\def \max{\text{max}}
\def \eff{\text{eff}}
\def \rel {\text{rel}}
\def \GeV {\text{GeV}}

\graphicspath{{./plots/}}

\begin{document}

\qrcode[hyperlink,version=10,height=2.5cm]{https://inspirehep.net/record/1374222}
\\
\vspace{2.5cm}
\title{Real time evolution of non-Gaussian cumulants in the QCD critical regime}
\author{Swagato~Mukherjee}
\affiliation{Department of Physics,
Brookhaven National Laboratory, Upton, New York 11973-5000}
\author{Raju~Venugopalan}
\affiliation{Department of Physics,
Brookhaven National Laboratory, Upton, New York 11973-5000}
\author{Yi~Yin}
\affiliation{Department of Physics,
Brookhaven National Laboratory, Upton, New York 11973-5000}

\date{ \today}

\begin{abstract}
We derive a coupled set of equations that describe the non-equilibrium evolution of cumulants of critical fluctuations 
for space-time trajectories on the cross-over side of the QCD phase diagram. In particular, novel expressions are obtained for the 
non-equilibrium evolution of non-Gaussian Skewness and Kurtosis cumulants. Utilizing a simple model of the space-time evolution of a heavy-ion 
collision, we demonstrate that, depending on the relaxation rate of critical fluctuations, Skewness and Kurtosis can differ significantly in magnitude as well as in 
sign from equilibrium expectations. Memory effects are important and shown to persist even for trajectories that skirt the edge of the critical regime. 
We use phenomenologically motivated parameterizations of freeze-out curves, and of the beam energy dependence of the net baryon chemical potential, to explore the 
implications of our model study for the critical point search in heavy-ion collisions. 

\end{abstract}
\maketitle

\section{Introduction}

The structure of the QCD phase diagram has attracted a large number of theoretical and experimental
studies~\cite{Stephanov:2004wx,Stephanov:2007fk,Fukushima:2010bq}. Of high interest is the possible existence of a conjectured 
critical point~\cite{Halasz:1998qr,Berges:1998rc} in the phase diagram. This critical point is the end
point of a first-order phase transition line that separates, in the chiral limit, a chirally symmetric quark-gluon plasma (QGP) phase from 
a hadron matter phase of QCD.  An entire experimental program, the Beam Energy Scan (BES) at the Relativistic
Heavy-Ion Collider (RHIC), is dedicated to searches for the QCD critical point~\cite{STAR-wp,Heinz:2015tua}.  

A universal feature of a system near the critical point is the growth and
divergence of the fluctuations of the order parameter. These fluctuations can be
quantified by the variance  of the critical field (the Gaussian cumulant), as well as higher non-Gaussian cumulants
 such as  skewness and kurtosis, which correspond, respectively, to the third and fourth cumulants.  They scale with the correlation length $\xi$, which is large near the
critical regime and divergent at the critical point.  

As pointed out in Ref.~\cite{Stephanov:2008qz}, the non-Gaussian cumulants are much more sensitive to
$\xi$ than the variance. For example, while the variance grows as $\xi^2$, the kurtosis grows far more rapidly as $\xi^{7}$. Further, even qualitative features of the non-Gaussian cumulants,  such as a change in sign and the associated non-monotonicity, can signal the presence
of criticality in the QCD phase diagram~\cite{Asakawa:2009aj,Friman:2011pf,Stephanov:2011pb}. As observed
in Ref.~\cite{Stephanov:2011pb}, the sign of kurtosis in equilibrium is negative when the critical
point is approached from the crossover side and positive when approached from the first
order transition side. 

These enhanced near critical fluctuations are  accessible through measurements of event-by-event fluctuations of various conserved quantities in
heavy-ion collisions~\cite{Stephanov:1999zu,Stephanov:2011pb,Jeon:2003gk} as well as fluctuations of particle
multiplicities~\cite{Hatta:2003wn,Athanasiou:2010kw}. As the QCD
phase diagram is scanned by varying the beam energy, non-Gaussian event-by-event fluctuations of multiplicities are generically expected
to show non-monotonic behavior in the proximity of a critical point. Uncovering such behavior, and cleanly identifying this behavior as a signature of criticality, is a major focus of the RHIC BES program~\cite{Aggarwal:2010wy,Aggarwal:2010wy,Adamczyk:2014fia,Adamczyk:2014fia,Luo:2015ewa,Mitchell:2012mx}.

The above expectations are entirely based on the assumption
that the soft modes responsible for critical fluctuations are in
equilibrium with the medium. Indeed if this were the case, fluctuations of
conserved quantities measured in heavy-ion experiments could be directly compared to equilibrium lattice QCD calculations~\cite{Bazavov:2012vg,Mukherjee:2013lsa,Bazavov:2014xya,Borsanyi:2014ewa}.
However the expanding medium created in heavy-ion collisions  only
spends a limited amount of time  in the QCD
critical regime and it is unlikely the critical modes 
remain in equilibrium in this duration.  
Therefore the relevant cumulants
 can in principle differ considerably from their equilibrium
values~\cite{Stephanov:1999zu,Berdnikov:1999ph}.
For instance, the time it takes the correlation length to reach its equilibrium value 
scales as $\tau_{\rm eff} \sim \xi^{z}$, 
which defines the dynamic scaling exponent $z$~\cite{RevModPhys.49.435,Son:2004iv}. 

Since the non-Gaussian cumulants
grow with higher powers of the correlation length, their relaxation times to their equilibrium values may be significantly larger. Even the
sign of a non-Gaussian cumulant may differ from its equilibrium value due to this critical slowing
down of relaxation rates.  These simple considerations suggest that non-equilibrium memory effects may play an important role in 
interpreting the results of BES measurements, and conversely, in applying lattice results to predict detailed features of the data. 

A number of papers have previously investigated critical dynamics off-equilibrium. The quasi-stationary dynamics of Gaussian fluctuations near the critical
point was studied in Ref.~\cite{Berdnikov:1999ph} by using a model equation
to compute the relaxation of the correlation length.  
Gaussian fluctuations in presence of the critical point were also
investigated within a hydrodynamic approach~\cite{Nonaka:2004pg}. Further, the hydrodynamic evolution of a quark-gluon fluid that is 
coupled to the classical real-time evolution of long wavelength modes of chiral
fields has been explored in a number of effective models that contain a critical point~\cite{Paech:2005cx,Nahrgang:2011mv,Herold:2013bi}. These
models have also been employed to investigate the out-of-equilibrium evolution of the relevant correlation length.  
Another class of models employs a coupled Boltzmann-Langevin--Vlasov kinetic approach, with conserved charges to model the evolution of critical fluctuations through chemical and kinetic freeze out~\cite{Stephanov:2009ra}. These can be contrasted~\cite{Schuster:2009jv} with results from UrQMD, a hadronic event generator that does not contain critical fluctuations, to isolate genuine effects due to critical fluctuations. All the studies outlined focus on the evolution of Gaussian cumulants. 
(See Refs.~\cite{Kitazawa:2013bta,Kitazawa:2015ira} for the evolution of higher order cumulants in a system without critical phenomena.)
Analyses incorporating critical fluctuations of non-Gaussian cumulants off-equilibrium are sorely lacking.

This paper is a first attempt to address this gap in our knowledge of non-Gaussian fluctuations and to understand its consequences. In particular, we compute the real-time evolution of
non-Gaussian cumulants of the critical field to estimate memory effects in the vicinity of the QCD critical point. The expressions for the cumulants are derived from the Fokker-Planck equation determining the evolution of the non-equilibrium probability distribution
function of the critical field $\s$. One obtains a set of coupled first order differential equations which
describe the evolution of the mean, variance, skewness and kurtosis, on the cross-over side of the QCD critical scaling regime, for trajectories that are close to but not at
the critical point. This coupled set of equations takes into account universal
equilibrium properties and non-equilibrium dynamics in the QCD critical regime and is the appropriate theoretical framework to study the
evolution of non-equilibrium cumulants in that regime.

The QCD critical point lies in the static universality class of the
3-dimensional Ising model~\cite{Berges:1998rc} for which the equilibrium cumulants are
well known~\cite{ZinnJustin:1999bf}. 
The latter can therefore be used to fix the equilibrium values of the cumulants of the QCD critical field. With these fixed, the non-equilibrium values of the cumulants can be determined from our evolution equations. A key quantity determining the non-equilibrium evolution of cumulants is the effective relaxation time $\tau_{\rm eff}$ of critical fluctuations. This quantity is unknown and can only be estimated to be a strong interaction time scale. For reasonable variations in its value, we will show that strong memory effects are seen for the non-Gaussian cumulants. 

While our expressions in terms of Ising variables are universal, the
relation of these to the temperature $T$ and the chemical potential
$\mu$ in QCD is non-universal, and is a source of significant
systematic uncertainty. Nevertheless, with physically motivated
assumptions, one can obtain expressions for the cumulants as a
function of temperature and chemical potential. There is a significant
amount of phenomenological information from thermal model ratios of
particle multiplicities that allows one to extract the chemical freeze-out curve in
temperature and chemical potential in the QCD phase diagram. Further,
the particle ratios, with model assumptions, can be used to relate
$\mu$ to the center of mass energy $\sqrt{s}$ of the heavy ion
collision. With these phenomenological inputs, our results can be used to provide qualitative
estimates of the importance of memory effects for the BES. 

We will demonstrate that, depending on the trajectory
followed, the sign of Skewness and Kurtosis can flip relative
to their equilibrium values. These results suggest the need for caution in interpreting, on the basis of equilibrium expectations, the results of experiments. We also note that if the system were in equilibrium throughout its evolution in the $T$-$\mu$ plane, it would have to pass very close to critical point to be sensitive to critical fluctuations. On the other hand, if memory effects are important, even a trajectory some distance away from the critical point may be sensitive to it.  

The paper is organized as follows.  In Sec.~\ref{sec:derive_eq}, we review the equilibrium 
properties of the  $\s$ field in the critical regime and
derive the non-equilibrium evolution equations for cumulants.  Our results up to this point are completely general and apply equally to all systems that lie in the 3-D Ising universality class. 
In Sec.~\ref{sec:universal-param}, we map our results to the QCD critical regime. Because the QCD critical point lies in the universality class of the 3-D Ising model, we will be able to specify the equilibrium properties that are necessary inputs for applying our equations to study the evolution of
cumulants in the QCD critical regime. We construct a simple model of the space-time evolution of a heavy-ion collision in Sec.~\ref{sec:space-time}. In Sec.~\ref{sec:traja}, we solve the evolution equations and present results for a representative trajectory passing through the critical regime.  We
further extend our formalism to model  the beam energy scan in Sec.~\ref{sec:BES}.  We summarize our results and discuss their implications for the QCD critical
point search in Sec.~\ref{sec:summary}.

\section{
\label{sec:derive_eq}
Evolution equations for cumulants}

\subsection{Critical fluctuations in equilibrium}
We  begin by considering the critical field $\s(\vx)$ and 
its zero momentum mode
\be
\s
\equiv \frac{1}{V}\int d^3x\, \s(\vx)\, ,
\ee
where $V$ is the volume.
The fluctuations of this zero-mode $\s$-field are described by the
probability distribution $P(\s;\tau)$ which in general depends on the (proper) time
$\tau$. 
Introducing the average of any $\s$-dependent functions with respect
to $P(\s;\tau)$,
\be
\label{eq:average_define}
\<\ldots\>
= \frac{\int^{\infty}_{-\infty} d\s\,
  \(\ldots\)P(\s;\tau)}{\int^{\infty}_{-\infty} d\s\, P(\s;\tau)}\, ,
\ee
one can define the expectation value of the $\s$-field $M$(the ``magnetization'') and the cumulants of the $\s$-field as
\begin{align}
\label{eq:cumulants_def}
M(\tau)
\equiv \<\s\>\, ,
\qquad
 \delta\s
\equiv
\s-M(\tau)
\nonumber
\\
\k_{2}(\tau)
\equiv \<\(\delta \s\)^{2}\>\, ,
\qquad
\k_{3}(\tau)
\equiv\< \(\delta\s\)^{3}\>\, ,
\nonumber
\\
\k_{4}(\tau)
\equiv \<
\(\delta\s\)^{4}\> -3\[\k_{2}(\tau)\]^2\, .
\end{align}

Before discussing the evolution of non-equilibrium critical fluctuations,
it is useful to first review their properties in equilibrium.
For the static universality class of the 
3-D Ising model,  the dependence of the equilibrium cumulants $\k^{\equ}_{n}$ on the Ising
variables (the reduced
temperature $r= (t-t_c)/t_c$, where $t$ is the Ising temperature,
$t_c$ the Ising critical temperature and $h$ the rescaled magnetic
field) is universal. 
The explicit expressions for these are given in Appendix~\ref{sec:EOS}. 

Alternatively, the critical fluctuations in equilibrium can be described by the distribution 
$P_{0}(\s)\sim \exp\(-{\cal E}_{0}(\s)/T\)$, where $T$ is the temperature.
Here we
expand the effective action functional ${\cal E}_{0}(\s)$ around the minimum
$\s_{0}$ in gradients and powers of $\s(\vx)$ as
\begin{align}
\label{eq:O0}
{\cal E}_{0}(\s)
= \int
d^3x
\Bigg\{\,
\frac{1}{2}\[\nabla\s(\vx)\]^2+\frac{1}{2}m^2_{\s}\[\s(\vx)-\s_{0}\]^2+
\nonumber
\\
\frac{\l_{3}}{3}\[\s(\vx)-\s_{0}\]^3+\frac{\l_{4}}{4}\[\s(\vx)-\s_{0}\]^4
\Bigg\}
\, ,
\end{align}
following Refs.~\cite{Stephanov:2008qz,Stephanov:2011pb}.
Near a critical point,
the mass ($m_{\s}$) of the $\s$ field, as well as the other parameters in \eq\eqref{eq:O0}, scale with the equilibrium value of the correlation length $\xi_{\equ}$
as~\cite{Stephanov:2008qz}\footnote{Following Refs.~\cite{Stephanov:2008qz,Stephanov:2011pb}, we 
shall neglect the anomalous scaling dimension $\eta$, which is only of order few percent, in this work.} 
\begin{align}
\label{eq:l_scaling}
m^{-1}_{\s}=\xi_{\rm eq}\, ,
\qquad
\s_{0}= \tsigma_{0}T (T\xi_{\equ})^{-1/2}\, , 
\nonumber
\\
\l_{3}= \tl_{3} T(T\xi_{\equ})^{-3/2}\, ,
\qquad
\l_{4}=\tl_{4}(T\xi_{\equ})^{-1}\, .
\end{align}
Expressed thus, the dimensionless parameters $\tsigma_{0},\tl_{3},\tl_{4}$ do not depend on $\xi_{\equ}$ and therefore remain finite when approaching the critical point. 
Because in this work we are interested in the fluctuations of the zero momentum mode of the $\s$
field, we will neglect the spatial dependence of $\s(\vx)$.
Consequently, the equilibrium distribution becomes
\be
\label{eq:P0}
P_{0}(\s)\sim\exp\(-V_{4}\O_{0}(\s)\)
\, , 
\qquad
 V_{4}\equiv\frac{ V}{T} \, ,
\ee
where $\O_{0}(\s)$ is a function of the zero mode of the $\s$-field,
\be
\label{eq:O_mean}
\O_{0}(\sigma)
=
\frac{1}{2}m^2_{\s} (\sigma-\sigma_{0})^2
+ \frac{\l_{3}}{3}(\sigma-\sigma_{0})^3
+ \frac{\l_{4}}{4}
(\sigma-\sigma_{0})^4 \, .
\ee

An  important quantity characterizing the probability distribution
\eq\eqref{eq:P0} is the
ratio between the correlation length and the size of the system,
\be
\label{eq:e_def}
\e =\sqrt{\frac{\xi^{3}_{\equ}}{V}}\, .
\ee
Throughout this work,
we will work in the scaling regime (near, but not at the critical point)
where the correlation length $\xi$ is
larger than any microscopic scale $L_{\rm micr}$ but smaller than the size of
the system, $L_{\rm micr}\ll \xi \ll L$. Hence $\e\ll 1$. 
In this regime, the kinetic term in \eq\eqref{eq:O0} is proportional to
$\s^2/L^2$ and is small compared to the mass term $\s^2/\xi^2_{\equ}$. This justifies our dropping the kinetic term in \eq\eqref{eq:O0}. 
Further, 
truncating the expansion at the $\l_{4}$ term in \eq\eqref{eq:O0}
is justified as well because higher order terms are suppressed for small $\e$. 

Treating $\e$ as an expansion parameter enables us to relate the parameters  $\s_{0},m_{\s}, \l_{3},\l_{4}$ to the equilibrium
``magnetization'' $M^{\equ}$ and the 
cumulants $\k^{\equ}_{n}, n=2,3,4$. To leading power in $\e$, we obtain, 
\begin{align}
\label{eq:eq_param}
M^{\equ}=\s_{0}
\, ,
\qquad
\k^{\equ}_{2}
= \frac{\xi^2_{\equ}}{V_{4}}
\, , 
\qquad
\nonumber
\k^{\equ}_{3}= -\frac{2\l_{3}}{V^2_4}\xi^{6}_{\equ}\, ,
\\
\k^{\equ}_{4}= \frac{6}{V^3_4}\[2(\l_{3}\xi_{\equ})^2-\l_{4}\]\xi^8_{\equ}\, ,
\end{align}
where we have employed \eq\eqref{eq:average_define}, \eqref{eq:cumulants_def} and \eqref{eq:O_mean}.  This result is in agreement with Ref.~\cite{Stephanov:2008qz}.
Using \eq\eqref{eq:l_scaling},
we also recover the scaling behavior $M^{\equ}\sim \xi^{-1/2}_{\equ},
\k^{\equ}_{2}\sim \xi^{2}_{\equ},\k^{\equ}_{3}\sim \xi^{9/2}_{\equ},
\k^{\equ}_{4}\sim \xi^{7}_{\equ}$.

\subsection{Evolution equation for cumulants}

Now that we have established the behavior of the equilibrium cumulants in our approach, we will turn to their non-equilibrium evolution with 
proper time $\tau$. The non-equilibrium transport properties of fluids near a critical point depend on their dynamical universality class; these were formalized in the 
classification scheme of Ref.~\cite{RevModPhys.49.435}. In the QCD case, which according to 
\cite{Son:2004iv,Fujii:2003bz,Fujii:2004za,Fujii:2004jt} should lie in the model H universality class of \cite{RevModPhys.49.435}, the relevant fields are those of the chiral order parameter (the chiral condensate) and the baryon density. 

It was shown in \cite{Son:2004iv} that the relevant hydrodynamical equations for the space-time evolution of these two fields can be expressed as a coupled set of Langevin equations. In the late time, long wavelength hydrodynamic asymptotic limit, the corresponding eigenvalue problem can be solved, revealing only one long wavelength diffusive mode, with a diffusion constant that goes to zero at the critical point. As argued in Ref.~\cite{Son:2004iv}
, both the chiral condensate and the baryon density fluctuate on these hydrodynamic time scales, with the chiral condensate seen as ``tracing" the profile of the baryon density as it relaxes to its equilibrium value. 

The diffusive properties of this critical mode are captured by the Fokker-Planck equation for the relaxation to equilibrium of the distribution $P(\sigma,\tau)$, expressed as 
\begin{multline}
\label{eq:FP1}
\pd_{\t}P(\sigma;\t)=\frac{1}{\(m^2_{\sigma}\tau_{\eff}\)}
\Bigg\{
\pd_{\sigma}
\[\pd_{\sigma}\O_{0}(\sigma)+V_{4}^{-1}\pd_{\sigma}\]P(\sigma;\t)
\Bigg\}
\, ,
\end{multline}
where $\tau_{\eff}$ is the effective relaxation rate. As we shall discuss further shortly, $\tau_{\eff}$ scales with a universal power of the correlation length $\xi$ that is characteristic of model H.

In a static medium where $T$ and $V$ are independent of time,
it is easy to check that \eq\eqref{eq:P0} is the static solution to
\eq\eqref{eq:FP1}.
For an expanding medium, the solution is more involved. Now $\O_{0}(\sigma)$ also depends on time
since the thermodynamical variables such as the temperature and the chemical
potential 
change with time.
Nevertheless, as long as the expansion rate does not exceed the characteristic time scales in the system, 
\eq\eqref{eq:FP1} will describe the evolution of the distribution towards a quasi-stationary fixed point solution.

We shall now derive the evolution equations for the cumulants.
We first note that
for any function $g(\sigma)$ which does not grow exponentially in the large $\sigma$ limit,
the evolution of  the expectation value $\<g(\sigma)\>$ is given by
\begin{eqnarray}
\label{eq:g_evo}
\pd_{\tau}\<g(\sigma)\>
= \int^{\infty}_{-\infty} d\sigma g(\sigma) \pd_{\tau}P(\sigma;\tau)
\nonumber
\\
=\frac{1}{\(m^2_{\s}\tau_{\eff}\)}\Bigg\{
\int^{\infty}_{-\infty} d\sigma g(\sigma)\[\O'_{0}(\sigma) P(\sigma;\tau)\]'
\nonumber
\\
+V_{4}^{-1}\int^{\infty}_{-\infty} d\sigma g(\sigma) P''(\sigma;\tau) \Bigg\}
\nonumber
\\
=- \frac{1}{\(m^2_{\s}\tau_{\eff}\)}\[  \<g'(\sigma)\O'_{0}(\s)\>-\frac{\<g''(\sigma)\>}{V_{4}}\]\, , 
\end{eqnarray}
where we have used \eq\eqref{eq:FP1} and the
definition \eq\eqref{eq:average_define}.
Here and hereafter,
we will use the prime symbol to denote the derivative with respect to $\s$.
To obtain the last line in the equality above, we performed an integration by parts.

We will first study the evolution of $M$ by taking
$g(\sigma)=\sigma$. From \eq\eqref{eq:g_evo}, 
we immediately obtain 
\be
\label{eq:kappa1_evo10}
\pd_{\tau} M(\tau)
=-\frac{1}{\(m^2_{\s}\tau_{\eff}\)}\<\O'_{0}(\sigma)\>
\, ,
\ee
where we have used \eq\eqref{eq:cumulants_def}. 
To express the R.H.S of \eq\eqref{eq:kappa1_evo10} in terms of cumulants, 
we first Taylor expand $\O'_{0}(\sigma)$ as a function
of $\sigma$ around $\s=M$:
\begin{eqnarray}
\label{eq:dOs}
\O'_{0}(\sigma)&=&
 \O'_{0}(M) +\O''_{0}(M)\delta\s
+\frac{\O'''_{0}(M)}{2!}\delta\s^2 \nonumber \\
&+&\frac{\O''''_{0}(M)}{3!}\delta\s^3 \, . 
\end{eqnarray}
As $\O_{0}(\sigma)$ is a polynomial of $\sigma$,
the above expansion is exact. 
Substituting \eq\eqref{eq:dOs} into \eq\eqref{eq:kappa1_evo10} and
using \eq\eqref{eq:cumulants_def},
we obtain, 
\begin{eqnarray}
\label{eq:kappa1_evo1temp}
\pd_{\tau} M(\tau)
&=&-\frac{1}{\(m^2_{\s}\tau_{\eff}\)}
\Bigg\{\O'_{0}(M) 
+\frac{\O'''_{0}(M)}{2!}\k_{2}(\tau)\nonumber \\
&+&\frac{\O''''_{0}(M)}{3!}\k_{3}(\tau)\Bigg\}\, .
\end{eqnarray}
The evolution equation \eq\eqref{eq:kappa1_evo1temp} up to this point is exact but not closed as the R.H.S of
\eq\eqref{eq:kappa1_evo1temp} depends on the non-equilibrium values of
$\k_{2},\k_{3}$. We shall now demonstrate that the 
the evolution of $\k_{2},\k_{3}$  decouples from the evolution of
$M(\tau)$ for small values of $\e$.

In order to arrive at this result, we first introduce the dimensionless functions
\begin{align}
\label{eq:Fs_def}
F_{n}(M)
\equiv  V_{4} \[\e^{2-n}b^{n}\pd^{n}_{\sigma}\O_{0}\(\s\)\]\bigg |_{\s=M} \, ,
\qquad
n=1, 2,3,4 \, ,
\end{align}
which, by construction, are finite in the small $\e$ limit. 
In \eq\eqref{eq:Fs_def}, the dimensionful quantity $b$ is the square root of the variance $\k^{\equ}_{2}$ 
in \eq\eqref{eq:eq_param},
\be
\label{eq:b_def}
b =\sqrt{\k^{\equ}_{2}}\equiv \sqrt{\frac{\xi^{2}_{\equ}}{V_4}}
\, .
\ee
The reader should keep in mind that that, for an expanding medium, $b$ will depend on $\tau$ due to the change of both $V$ and
$\xi_{\equ}$ with proper time. Explicitly working out \eq\eqref{eq:Fs_def},
we obtain
\begin{eqnarray}
\label{eq:F_exp}
F_{1}(M)
&=& \d \tM\[1+ \tl_{3} (\d \tM)+\tl_{4}(\d \tM)^2\] \, ,
\nonumber
\\ 
F_{2}(M)
&=&1+ 2 \tl_{3}(\d \tM)+3\tl_{4}(\d \tM)^2\, ,
\nonumber\\
F_{3}(M) 
&=&2 \[\tl_{3}+3\tl_{4} (\d \tM)\]\, ,
\,\,\,\,
 F_{4}
=6\,\tl_{4}\, .
\end{eqnarray}
On the R.H.S, we have defined, for convenience, the dimensionless quantity $\delta \tM \equiv \e
\(M-\sigma_{0}\)/b$.

Using \eq\eqref{eq:cumulants_def},  the evolution equation for $M(\tau)$ can be expressed as 
\begin{eqnarray}
\label{eq:kappa1_evo1}
 \pd_{\tau}M(\tau)
& &= -\tau^{-1}_{\eff}\(\frac{b}{\e}\) \Bigg\{\, F_{1}(M)\nonumber\\
&+& \frac{ \e^2}{2}\(\frac{\k_{2}}{b^{2}}\)F_{3}(M)
+\frac{\e^4}{6}\(\frac{\k_{3}}{\e\, b^{3}}\)F_{4}\,
\Bigg\}\, .
\end{eqnarray}

We shall now consider \eq\eqref{eq:kappa1_evo1} in the small $\e$ limit.
In powers of $\e$,
we will count 
\be
\label{eq:moments_counting}
\frac{\e M}{b}
\, ,  \,
 \frac{\k_{2}}{b^{2}} 
 \, ,  \,
 \frac{\k_{3}}{\e b^{3}} 
 \, ,  \,
 \frac{\k_{4}}{\e^{2}b^{4}} 
 \, ,  \,
\sim {\cal O}(1)\, . 
\ee
This power counting, as is clear from \eq\eqref{eq:eq_param}, holds for the equilibrium cumulants. 
In the following, we will assume that \eq\eqref{eq:moments_counting} also holds for the 
non-equilibrium cumulants. To the extent that the Fokker-Planck ``master" equation is valid, this assumption is reasonable. 
We will confirm later that this power counting is consistent with our
evolution equations at all times. 
Consequently,
the $\k_{2}$ and $\k_{3}$ terms on the R.H.S of \eq\eqref{eq:kappa1_evo1} are
suppressed  by $\e^2$  and $\e^{4}$ respectively and 
one obtains the closed form expression for $M$ to be
\bes
\label{eq:kappa_evo}
\begin{eqnarray}
\label{eq:kappa1_evo}
\pd_{\tau}M(\tau)
&=& -
\tau^{-1}_{\eff}\(\frac{ b}{\e}\)F_1(M)\[1+{\cal O}(\e^2)\]\,.
\end{eqnarray}
We wish to emphasize again that while $b, \e$ is independent of time for a static medium, for the expanding medium we will 
study in this paper  $\e, b$ will be time dependent due to the change of volume $V(\tau)$,
temperature $T(\tau)$ and $\xi_{\equ}$.

The above derivation can be straightforwardly extended to obtain evolution equations for
higher cumulants.
The details are given in Appendix.~\ref{sec:derivation_A}.
In general, the evolution of $\k_{n}$ is coupled to both lower
cumulants $\k_{n-1},\k_{n-2}$ and higher cumulants $\k_{n+1},\k_{n+2}$. 
However the coupling to higher cumulants is suppressed by powers of
$\e$. 
Keeping contributions at leading order in $\e$ from \eq\eqref{eq:kappa_evo_Full} in Appendix.~\ref{sec:derivation_A}, 
we obtain the expressions, 
\begin{eqnarray}
\pd_{\tau}\k_{2}(\tau)
&=&-2\,\tau^{-1}_{\eff}
\(b^2\)\[ \(\frac{\k_{2}}{b^2}\)F_{2}(M) -1\]
\[1+{\cal O}(\e^2)\]
\, ,
\nonumber\\
\pd_{\tau}\k_{3}(\tau)
&=&- 3\,\tau^{-1}_{\eff}\(\e \,b^{3}\)
\[ \( \frac{\k_{3}}{\e \,b^3}\)F_{2}(M) +
 \(\frac{\k_{2}}{b^2}\)^2F_3(M)\]
\nonumber\\
&\times&
\[1+{\cal O}(\e^2)\]
\, , 
\nonumber \\
\pd_{\tau}\k_{4}(\tau)
&=&-4\,\tau^{-1}_{\eff}\(\e^{2}\,b^{4}\)
\Bigg\{\, 
\(\frac{\k_{4}}{\e^2\,b^4}\) F_{2}(M)
\nonumber\\
&+&
3\(\frac{\k_{2}}{b^2}\)\(\frac{\k_{3}}{\e \,b^3}\) F_{3}(M)
+
\(\frac{\k_{2}}{b^2}\)^3 F_4\,
\Bigg\}\nonumber\\
&\times&
\[1+{\cal O}(\e^2)\]\, .
\end{eqnarray}
\ees
As the R.H.S of Eqs.~\eqref{eq:kappa_evo} only depends on $M,\k_{n},n=2,3,4$,
this system of equations is closed.
Eqs.~\eqref{eq:kappa_evo} are a key result of this paper, and to best of our knowledge, are new in the literature.
Employing the power counting in \eq\eqref{eq:moments_counting},
 one observes that the R.H.S of the evolution equations satisfies this power counting. Therefore if initially 
the power counting \eq\eqref{eq:moments_counting} is satisfied, it is preserved for all subsequent times. 
Since in our derivation we do not assume that the medium is static, Eqs.~\eqref{eq:kappa_evo} are well suited for studying dynamical systems such as those created in heavy-ion collisions.

\subsection{
\label{sec:Gaussian}
Two limiting cases
}
It is instructive to examine the evolution equations Eq.~\eqref{eq:kappa_evo} in  limiting cases.
We first consider the limit where the equilibrium probability distribution
$P_{0}(\s)$ is the Gaussian distribution,
\be
{\O}_{0}(\s)
= \frac{1}{2}m^2_{\s}\(\sigma -\sigma_{0}\)^2\, .
\ee
For this case, the equilibrium values of $\kappa_{n}$ are simply, 
\be
M^{\equ}=\s_{0}\, ,
\qquad
\k^{\rm eq}_{2}= b^2\, ,
\qquad
\k^{\equ}_{3}=\k^{\equ}_{4}=0\, .
\ee
Further, 
$F_{1}=\delta \tM, F_{2}=1 , F_{3}=F_{4}=0 $, and 
Eqs.~\eqref{eq:kappa_evo} reduce to 
\be
\label{eq:gaussian_evo}
\pd_{\tau}\k_{n}
=-n\,\tau^{-1}_{\eff}\[\k_{n}(\tau)-\k^{\equ}_{n}\]\, , 
\qquad
n=1,\ldots
\ee
where note that $\k_1$ is shorthand for the magnetization $M$. 
\eq\eqref{eq:gaussian_evo} expresses the fact that if the equilibrium probability
distribution of the $\s$-field is a Gaussian, the evolution of cumulants are decoupled.
For $\k_{n}$, the damping rate is proportional to $n$; hence the higher cumulants approach their equilibrium values earlier than lower cumulants.

With certain assumptions, the results in this Gaussian limit can be
shown to be identical to those obtained previously in
Ref.~\cite{Berdnikov:1999ph}. The latter follow from the rate equation conjectured to describe the evolution of
the non-equilibrium correlation length, 
\be
\label{eq:BR}
\pd_{\tau}\[\xi^{-1}(\tau)\]
= -\tau^{-1}_{\eff}
\[\xi^{-1}(\tau)-\xi^{-1}_{\equ}(\tau)\]\, . 
\ee
To facilitate a comparison of this equation to \eq\eqref{eq:gaussian_evo}, one identifies the non-equilibrium correlation length to be 
\be
\label{eq:xi_eff}
\xi(\tau)\equiv
\sqrt{V_4\,\k_{2}(\tau)}\, .
\ee
Eq.~\eqref{eq:gaussian_evo}, for $n=2$, can then be expressed as 
\be
\label{eq:xi_evo}
\pd_{\tau}\[\frac{\xi^2(\tau)}{V_{4}}\]
= -2\, \tau^{-1}_{\eff}
\[ \frac{\xi^2(\tau)}{V_{4}}- \frac{\xi^2_{\equ}}{V_{4}}\]\, . 
\ee
If we require the medium to be static and require further that deviations of $\xi(\tau)$ from the
equilibrium value $\xi_{\equ}$ are small, we then find that 
\eq\eqref{eq:xi_evo} reduces to \eq\eqref{eq:BR} after a rescaling of $\tau_{\eff}$ by a factor of 2.

Another interesting limit is the near equilibrium limit where $\delta \k_{n}=\k_{n}-\k^{\equ}_{n}$ 
becomes small.
In this case,
the evolution equations can be linearized to read, 
\begin{eqnarray}
\label{eq:kappa_evoa}
 \pd_{\tau} M(\tau)
&=& -
\tau^{-1}_{\eff}\delta M
\, ,
\qquad
\pd_{\tau}\k_{2}(\tau)
=- 2\,\tau^{-1}_{\eff}\delta\tk_{2}
\, ,
\nonumber
\\
\pd_{\tau}\k_{3}(\tau)
&=&- 3\,\tau^{-1}_{\eff}(\e b^3)
\[ \(\frac{\delta \k_{3}}{\e b^3}\)+
4\tl_{3}\(\frac{\delta\k_{2}}{b^2}\)\]
\, ,
\nonumber
\\
\pd_{\tau}\[\k_{4}(\tau)\]
&=&-4\,\tau^{-1}_{\eff}(\e^2 b^4)
\Bigg\{\, 
\(\frac{\delta\k_{4}}{\e^2 b^4}\)+6\tl_{3}\(\frac{\delta\k_{3}}{\e b^3}\)
\nonumber
\\
&-&6\(2\tl^2_{3}-3\tl_{4}\)\(\frac{\delta\k_{2}}{b^2}\)
\Bigg\}\, .
\end{eqnarray}
From the power counting in \eq\eqref{eq:moments_counting},
all the terms in the $\[\ldots\]$ of the R.H.S of  \eq\eqref{eq:kappa_evoa} are
in the same order in power of $\e$. One can make the following observations about the evolution of cumulants in this particular limiting case. 
Firstly, unlike the Gaussian limit, the evolution of higher cumulants are coupled to the evolution of $\k_{2}$. 
It follows further, in contrast to the Gaussian limit,  that the higher cumulants will approach equilibrium only after $M$ and $\k_{2}$ reach their respective equilibrium values.

\subsection{A brief summary
\label{sec:non_Gaussian}}
We derived in this section a set of coupled first order differential equations, Eqs.~\eqref{eq:kappa_evo}, describing the evolution of the first four cumulants, $\k_{n}, n=1,2,3,4$ of the zero mode 
$\s$ of the critical field in the vicinity of the critical point. A key feature of these equations is that the evolution of higher moments are only coupled to the evolution of lower moments, and not vice versa. One therefore obtains a closed form system of equations that can be solved numerically. 

In general,  the evolution of the first four cumulants will be coupled to the evolution of higher cumulants, $\k_{n}$ including $n>4$ as well. 
However in the $\e \ll 1$ limit, where are results are strictly applicable, we demonstrated analytically that the coupling of lower cumulants to higher cumulants is suppressed.
Therefore the system of equations we derived is applicable for
describing the temporal evolution of moments in the scaling regime
$L_{\rm micr}<\xi_{\equ}<L$, where $\e \ll 1$ is satisfied.
For small values of $\e$, 
we have checked explicitly that the difference between computing the
non-equilibrium cumulants from numerical solutions of the
Fokker-Planck master equation and from solutions of the evolution
equations  Eqs.~\eqref{eq:kappa_evo} is suppressed by $\e$ 
(see Appendix.~\ref{sec:derivation_A} for further discussion of this point).

We could of course in principle have solved the master equation in \eq\eqref{eq:FP1} directly. However, this is not advisable for both practical and conceptual reasons. 
Firstly, from a practical perspective, numerically solving Eqs.~\eqref{eq:kappa_evo} is much faster than solving the Fokker-Planck equation. 
Furthermore, the interplay between cumulants is far more  transparent in the former approach. From a conceptual point of view, solving the Fokker-Planck equation \eq\eqref{eq:FP1} would \textit{not} give a more faithful representation of how cumulants evolve in the critical
regime. This is because an important input into \eq\eqref{eq:FP1} is the equilibrium distribution function $P_{0}(\s)$, which is not easy to obtain. From universality, this distribution is also the equilibrium distribution of the $3$-D Ising model. However, while the cumulants $\k^{\equ}_{n}$ in the $3$-D Ising model are known, reconstructing $P_{0}(\s)$ is non-trivial when $\e$ is not small. 

Thus conceptually there is no advantage 
in solving the Fokker-Planck equation at large $\e$ and no practical
advantage 
in solving it for small $\e$. 
Hence for the current state of the art, the derived Eqs.~\eqref{eq:kappa_evo} provide the most complete and consistent information, albeit limited, on the evolution of non-equilibrium moments in the critical region. We note further that as inhomogeneities may be important for bubble nucleation in first order transitions, our studies will be restricted to the 
cross-over side of the critical point. 

In this work, we shall concentrate on the evolution of the first four cumulants. It is straightforward to extend  Eqs.~\eqref{eq:kappa_evo}, if so desired, to include the 
evolution of even higher cumulants such as $\k_{5}$ and $\k_{6}$.
 
\section{Out of Equilibrium evolution of cumulants in the QCD critical regime
\label{sec:set_up}
}

In this section, we will apply Eqs.~\eqref{eq:kappa_evo} to study 
the evolution of cumulants in the critical scaling regime of QCD. Our discussion is organized as follows. We will first discuss the problem in the context of the 
3-D Ising model since it lies in the static universality class of the QCD critical point. This is important for determining the equilibrium values of the cumulants that provide the initial conditions for the evolution equations. In this context, we will also discuss the mapping of the Ising scaling regime to QCD as well as the dynamical universality class  governing the transport properties of the medium near the critical point. Next, for the purposes of our study, we will construct a simple model of the medium that mimics the expanding fireball formed in heavy ion collisions. Finally, within the framework of this simple model, we shall present our results for the temporal evolution of cumulants along a representative trajectory in the QCD critical scaling regime. 

\subsection{Fixing parameters from universality}
\label{sec:universal-param}

The inputs required to solve Eqs.~\eqref{eq:kappa_evo} along a trajectory passing through
the QCD scaling regime include $\s_{0},m_{\s},\l_{3},\l_{4}$ and the effective
relaxation time $\tau_{\eff}$. 
As we will also discuss shortly, 
we will need to know how $V, T$ change with $\tau$ for the case of an expanding medium. 

As noted previously, explicit expressions in the 3-D Ising model for the equilibrium $\k^{\equ}_{n}(r,h)$ as a function of the Ising variables are given in Appendix~\ref{sec:EOS}. 
With these expressions, we can determine the parameters $\s_{0}(r,h),m_{\s}(r,h),\l_{3}(r,h),\l_{4}(r,h)$ from 
\eq\eqref{eq:eq_param}. 
%
The dependence of $\tau_{\eff}$ on $\xi_{\equ}$ is universal and can be expressed in terms of  the dynamical critical exponent $z$ as 
\be
\label{eq:tau_eff1}
\tau_{\eff} = \tau_{\rel}
\(\frac{\xi_{\equ}}{\xi_{\min}}\)^{z}\, .
\ee
Here $\tau_{\rel}$ can be interpreted as the relaxation time at the outside edge of the critical region, defined by a minimal correlation length, 
$\xi=\xi_{\min}$. As we discussed earlier, the work of
Ref.~\cite{Son:2004iv} demonstrated that transport properties in the
vicinity the QCD critical point are governed by the diffusion of the
baryon density\footnote{It is also argued in Ref.~\cite{Son:2004iv} that the conservation (or not) of the isospin density does not influence considerations of universality because the isospin susceptibility, unlike the baryon susceptibility, is finite at the critical point.}. Because the chiral order parameter is not conserved and mixes with the baryon density, it does not influence the dynamical universality class. Consequently, the dynamical universality class is that of the liquid-gas phase transition, model H; in the classification scheme of Ref.~\cite{RevModPhys.49.435}, this gives $z=3$. It is clear from \eq\eqref{eq:tau_eff1} that the relaxation of the critical mode to equilibrium is greatly slowed down as a  spacetime trajectory in the system approaches the critical point. 

%
\begin{figure}
        \includegraphics[width=0.45\textwidth]{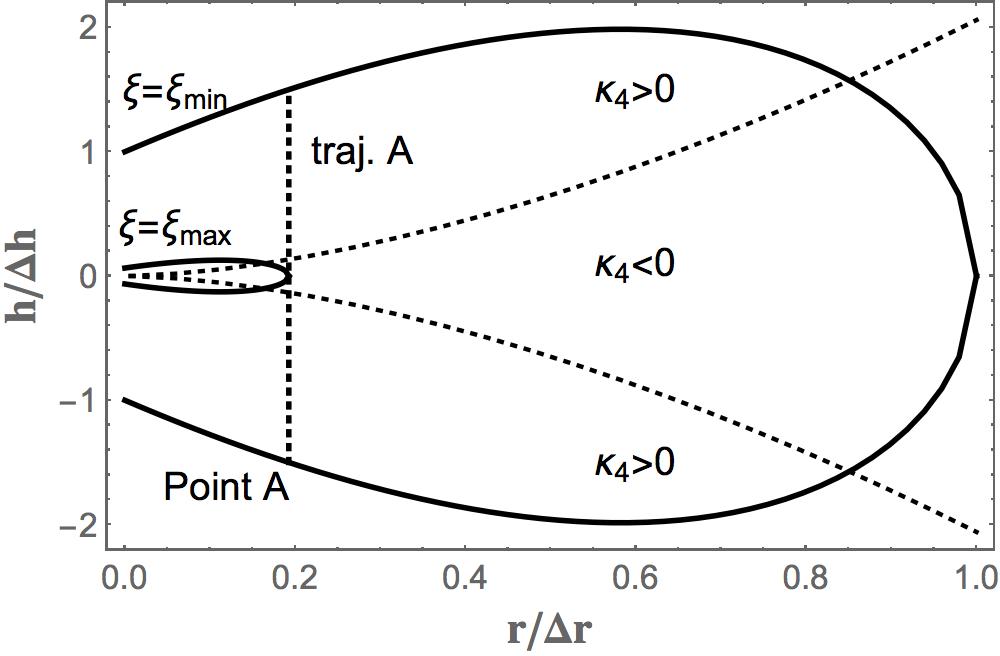}
              \caption{
      \label{fig:BES}
(Color online) A sketch of the cross-over side of the scaling regime in the $r, h$ plane.
We define the scaling regime with the criterion $\xi_{min}<\xi_{\rm
  eq}<\xi_{\max}$. To be specific, we take
$\xi_{\max}/\xi_{\min}=3$. The solid curve delineates the boundary of the critical regime.
The dotted curves show where the equilibrium value of the kurtosis
$K$ changes sign. The trajectory A  (see text) is shown with the black dashed vertical line. 
}
\end{figure}

Our discussion thus far only relied on the static and dynamical universal
properties of QCD critical point. 
However, as we are interested in the evolution of cumulants in the QCD critical regime,
we also need to specify the map\footnote{This map is \textit{non-universal} which is  a significant source of systematic uncertainty in quantitative dynamical studies 
in the vicinity the QCD critical point.} between the Ising variables $r, h$
to the QCD variables $T, \mu$. 
In the Ising model, the critical point is at $r= h=0$. 
There is a first order transition for $h=0,r<0$ and a cross-over for $h=0,r>0$.
Therefore the $r$-axis is the direction tangential to the line of first
order transition ending at $T_c$.  In contrast, in general, the h-axis will deform after the map to the $T, \mu$ plane. How this occurs is not known; 
for simplicity, we will follow the prescription of previous studies~\cite{Berdnikov:1999ph,Nonaka:2004pg} and assume that $h$ is perpendicular to the $r$-axis. Specifically, we 
will assume that $h$ is parallel to the $T$ axis in the QCD phase diagram. 
We then have following linear mapping relations,
\be
\label{eq:Tmu_mapping}
\frac{T- T_c}{\Delta T}
=\frac{h}{\Delta h}\, ,
\qquad
\frac{\mu-\mu_c}{\Delta \mu}
= -\frac{r}{\Delta r}\, ,
\ee
where $\Delta T, \Delta \mu$ denote the width of the critical regime in the QCD
phase diagram. The corresponding  width of the critical regime in the Ising variables $\Delta r,
\Delta h$ is defined to be
\be
\xi(r=\Delta r, h=0)=\xi(r=0,h=\Delta h)
= \xi_{\min}\, .
\ee
The inner and outer boundary of the critical regime in the $r,h$ plane are illustrated in
 Fig.~\ref{fig:BES}, where the curves are obtained from \eq\eqref{eq:kappa_rh} and \eq\eqref{eq:rh_Rtheta}, for $\xi_{\max}/\xi_{\min}=3$.
The equilibrium kurtosis is even with respect to $h$. Due to the potential importance of observables for  the sign of kurtosis,
 we also plot in Fig.~\ref{fig:BES} the boundary (from \eq\eqref{eq:kappa_rh} and \eq\eqref{eq:rh_Rtheta}) where the equilibrium kurtosis flips sign.
It is worth pointing out that, in contrast, the equilibrium skewness is an odd function of $h$.
One may check from Eqs.~\eqref{eq:rh_Rtheta}, \eqref{eq:th} and
  \eq\eqref{eq:kappa_rh}, that the skewness is positive for negative $h$. Given our choice of the direction of $h$-axis in
\eqref{eq:Tmu_mapping}, the equilibrium skewness is negative above the cross-over and is positive below it. 

\subsection{Simple model of space-time evolution in the vicinity of the critical point}
\label{sec:space-time}
For a given center of mass energy ($\sqrt{s}$), the space-time trajectory in the QCD phase diagram, 
can be determined by hydrodynamical simulations if the expansion rate is smaller than characteristic scattering rates.  
For the illustrative purposes of the study in this paper, a simple dynamical model of space-time evolution is sufficient. We will assume that for fixed $\sqrt{s}$,
the $\mu$ of the fireball is constant during the evolution.  
This trajectory corresponds  to the vertical dotted line in the
critical regime shown in Fig.~\ref{fig:BES}. It implicitly assumes of course the mapping relation \eq\eqref{eq:Tmu_mapping}. 

We parametrize the evolution of the volume as
\be
\label{eq:V_tau}
\frac{V(\tau)}{V_{I}}
= \(\frac{\tau}{\tau_I}\)^{n_{V}}\, ,  
\qquad
\frac{T(\tau)}{T_{I}}=\(\frac{\tau}{\tau_I}\)^{-n_V c^2_s}\, , 
\ee
where $V_{I}$ and $T_{I}$ are respectively the volume and temperature of the system at
$\tau_{I}$, the time at which the trajectory first hits the boundary of critical regime.
Here $n_V$ controls the rate of expansion; $n_V=3$ corresponds to a 3 dimensional Hubble-like expansion and
$n_V=1$, a 1 dimensional Bjorken-like expansion.
To obtain $T(\tau)$ in \eq\eqref{eq:V_tau},
we assumed that the total entropy is approximately conserved
during the evolution, and hence the entropy density goes as $s(\tau)\sim \tau^{-n_V}$.
We also used the relation $d\log T/d\log s=c^2_s$, where $c^{2}_s$ is the speed of sound.
Even at the highest heavy ion collision energies, the system is more
Hubble-like than Bjorken-like when trajectories approach the critical
point. We will therefore take $n_V=3$ for our study. Further, guided
by lattice measurements~\cite{Borsanyi:2012cr,Borsanyi:2013bia,Bazavov:2014pvz,Hegde:2014wga}
which indicate that $c^{2}_{s}$ is around $0.15$ near the QCD cross-over line, we will pick this value for the study in this paper. 

Finally, we need to specify the initial conditions for the solution of Eqs.~\eqref{eq:kappa_evo}. We will assume initially that 
$M, \k_{n},n=2,3,4$ are equal to their equilibrium values 
at $\tau=\tau_{I}$.
Choosing the initial volume is tricky, because, for the reasons outlined previously, we wish to ensure that $\e \ll 1$ at all times. We do this by 
requiring that the maximal value of $\e$ is imposed to be $\e_{c} \equiv  \sqrt{\frac{\xi^3_{\max}}{V_{c}}}
= 0.1$. Here $V_{c}$ is the volume when the trajectory followed by the system crosses $T_c$ in the cross-over region. 
As long as the change of the equilibrium correlation length is faster than the expansion of the volume, $\e_{c}$ gives the upper bound of $\e$ during the evolution of the
medium\footnote{For $\e\ll 1$, our results are independent
of the choice of $\e$.}. For instance, for $\xi_{max}=3$~fm, 
$\e_{c}=0.1$  corresponds to $V_{c}\approx (14~fm)^3$. We can then use the time evolution of the temperature in 
\eq\eqref{eq:V_tau} to determine $\tau_I$, and subsequently the corresponding equation for $V(\tau)$ to determine $V_I$ in terms of $\tau_I$. 

With these model assumptions, we are now ready to solve  Eqs.~\eqref{eq:kappa_evo} for each
given trajectory passing through the critical regime. 
Our results will only depend on one dimensionless parameter, 
$\tau_{\rel}/\tau_{I}$, where $\tau_{\rel}$ is the relaxation time at the boundary of critical regime. As there are no extant first principles calculations of $\tau_{\rel}$ in the 
QCD critical regime, we shall  take its value as a free parameter and solve \eq\eqref{eq:kappa_evo} for different choices of $\tau_{\rel}/\tau_{I}$. 
We can benchmark this value  by noting that if i) $\tau_{\rel}$ is 
$1$~fm, a characteristic strong interaction scale, and ii) $\tau_{I}$, 
the initial time at which the system enters the critical regime, is $10$~fm,
a reasonable estimate would be  $\tau_{\rel}/\tau_{I}\sim 0.1$.

\subsection{Results for cumulant evolution along a representative trajectory
\label{sec:traja}
}

For later convenience, we will express the evolution of the non-Gaussian cumulants in terms of the dimensionless quantities skewness ($S$) and kurtosis ($K$):
\be
\label{eq:SK_def}
S\equiv \[\frac{\tilde{V}^{1/2}_{4} \k_{3}}{\k^{3/2}_{2}}\]\, , 
\qquad
K \equiv \[\frac{\tilde{V}_{4}\k_{4}}{\k^{2}_{2}}\]\, , 
\ee
where $\tilde{V}_{4}$ is the rescaled $V_{4}$, 
\be
\tilde{V}_{4}= \frac{V/T}{V_{c}/T_c}\, .
\ee
With these definitions, we have deliberately taken the four-volume dependence out, as one should in 
explorations of critical behavior.  
 
\begin{figure}
\centering
       \includegraphics[width=0.5\textwidth]{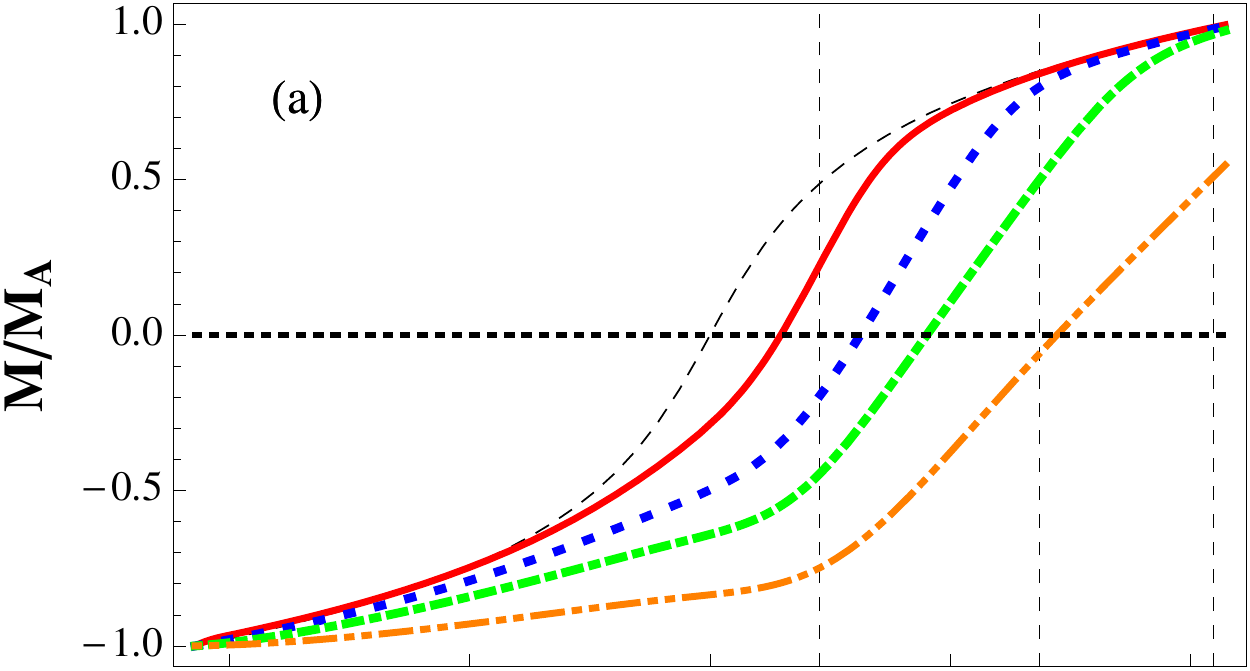}
       \includegraphics[width=0.5\textwidth]{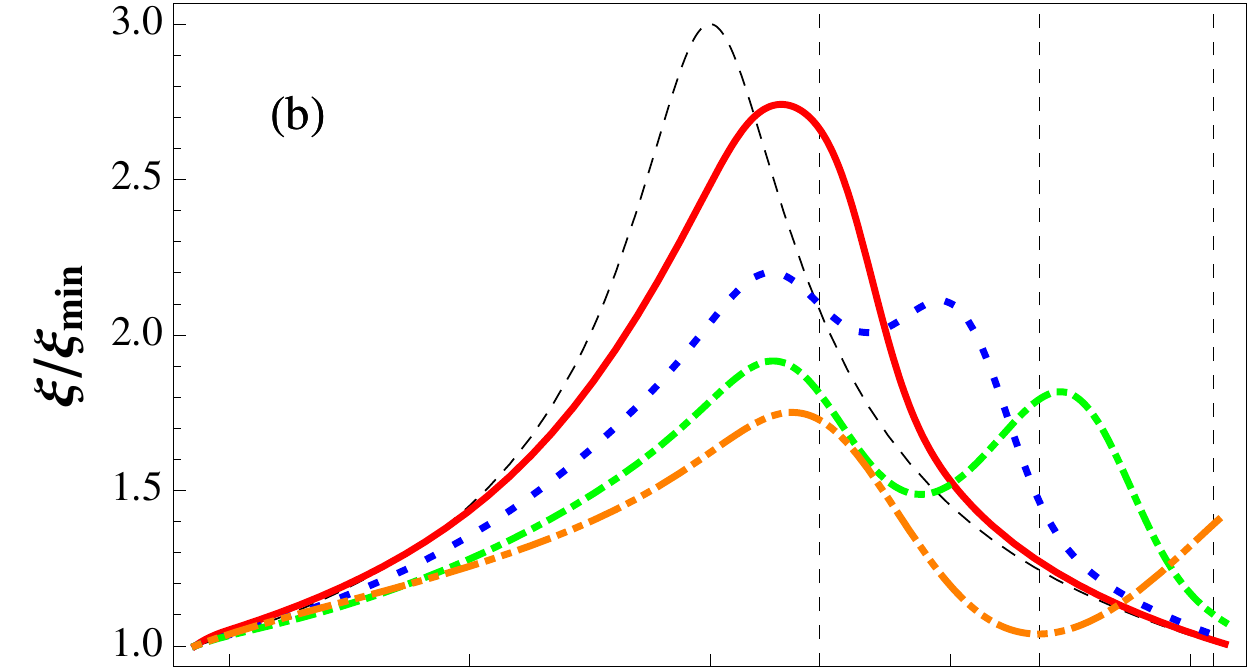}
   \includegraphics[width=0.5\textwidth]{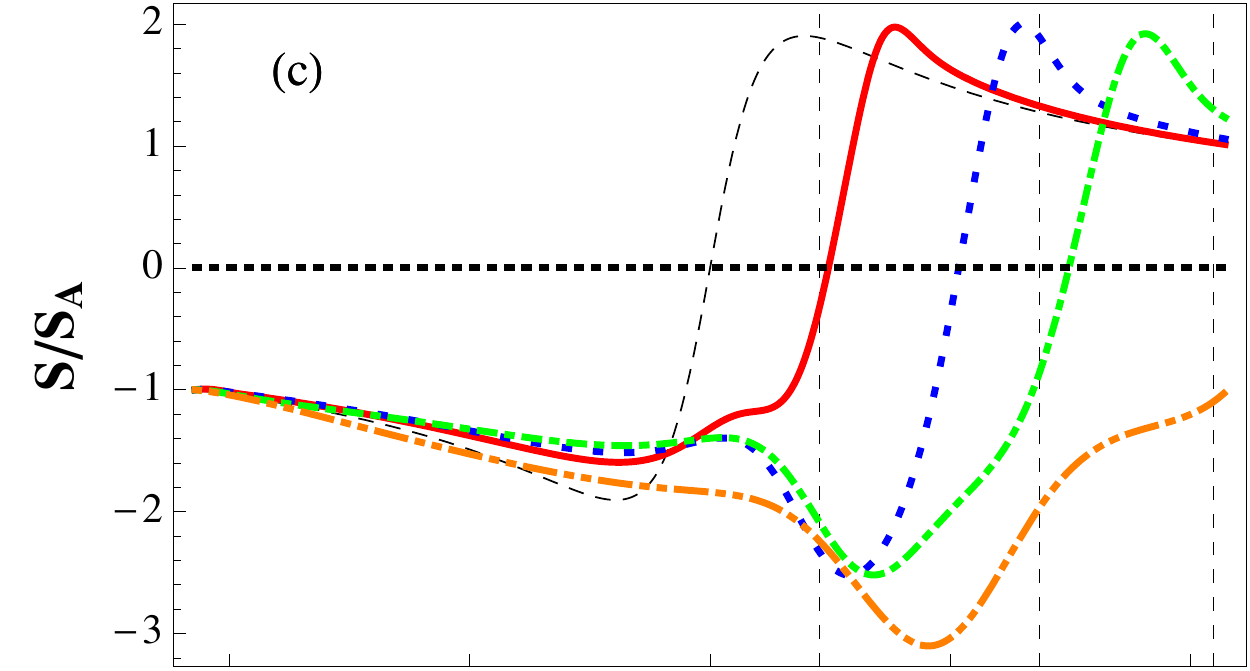}     
        \includegraphics[width=0.5\textwidth]{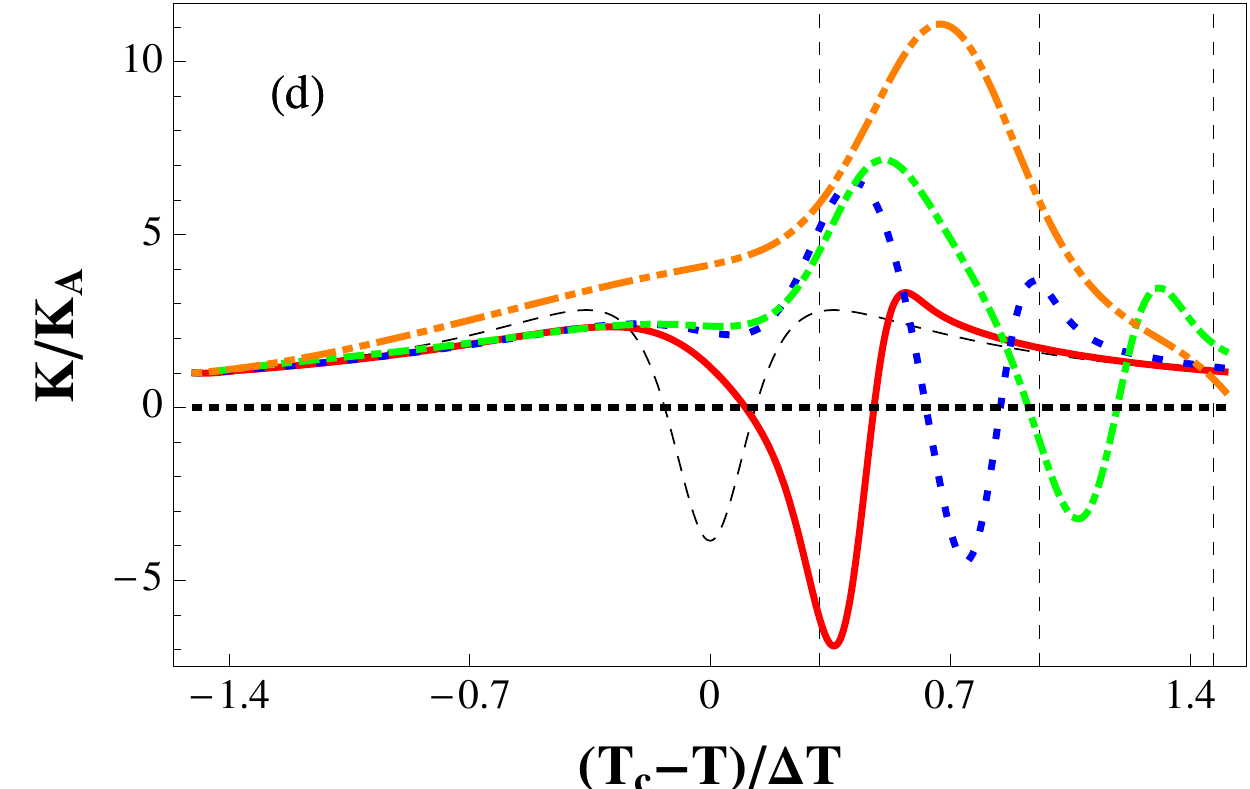}
\caption{
\label{fig:traja}(Color online)
(From top to bottom) The evolution of non-equilibrium mean
$M/M_{A}$(a), effective correlation length $\xi/\xi_{\rm min}$ (b), skewness $S/S_{A}$ (c) and kurtosis $K/K_{A}$ (d) 
as a function of $(T_{c}-T)/\Delta T$ along trajectory A~(c.f.~Fig.~\ref{fig:BES}).
Results for 
$\tau_{\rel}/\tau_I=0.005,0.02,0.05,0.2$ are shown in solid red, dotted blue, single dot dashed green, double dot dashed orange curves respectively. 
The dashed curves plot the corresponding equilibrium values. 
All results are normalized by the corresponding equilibrium value at
the end point trajectory A (c.f.~Fig.~\ref{fig:BES}).
The dashed vertical lines illustrate the $T$ at the point where the
a freeze-out curve intersects with trajectory A (from left to right, freeze-out curves of type I, II, III
respectively. For a discussion, see text in Sec.~\ref{sec:sqrts}.).
 }
\end{figure}

We will first study the evolution of the non-equilibrium cumulants along the 
representative trajectory A in Fig.~\ref{fig:BES}. Along this particular trajectory, 
$\xi_{\equ}$ will approach $\xi_{\rm max}$ when the trajectory approaches the
cross-over line. Memory effects are therefore most
prominent along trajectory A. For each fixed $\mu$,  trajectories can be parametrized by $h$, or equivalently $T$, via the mapping previously outlined. 
We plot in Fig.~\ref{fig:traja}, as a function of varying $T$, the non-equilibrium ratios $M/M_{A}, \xi/\xi_{\rm min}
,S/S_{A},K/K_{A}$. Here $M_{A}, \xi_{\rm min}, S_{A}, K_{A}$ are the equilibrium values specified at the end point of the trajectory. 

One immediately observes that non-equilibrium effects are important for all cumulants. 
The difference between the non-equilibrium cumulants and equilibrium
cumulants is apparent unless $\tau_{\rel}/\tau_{I}$ is much smaller than the noted benchmark value. (See for instance the red curves in
Fig.~\ref{fig:traja} that correspond to solutions with $\tau_{\rel}/\tau_{I}=0.005$.)
The difference is most visible near the cross-over line ($T=T_c$) where $\xi_{\equ}$
reaches its maximum value along the trajectory. This deviation is a direct manifestation of the effects of critical slowing down.

We shall now examine in detail the evolution for each individual non-equilibrium cumulant.
The behavior of $M(T)$ is relatively simple. 
As Fig.~\ref{fig:traja}~(a) shows, $M$ tends to follow the equilibrium values, though with critical slowing down, the change in sign occurs later for larger values of 
$\tau_{\rel}$.
$\xi_{\equ}$ is an even function of the Ising variable $h$--due to the
mapping relation \eq\eqref{eq:Tmu_mapping}--it is symmetric with respect to $T_{c}$.
Our results for the non-equilibrium correlation length are qualitatively similar to previous studies based on
 solving the rate equation \eq\eqref{eq:gaussian_evo}~\cite{Berdnikov:1999ph,Nonaka:2004pg}.
The role of memory effects on $\xi$ is two-fold. 
On the one hand, 
when the equilibrium correlation length $\xi_{\equ}$ is large, 
the effects of critical slowing down delay the growth of the 
effective non-equilibrium correlation length $\xi$. 
For example, as shown in Fig.~\ref{fig:traja}~(b), when $T$ is close to $T_{c}$ and $\xi_{\equ}$ approaches its maximum, 
the non-equilibrium $\xi$ for all $\tau_{\rel}$ under consideration are  smaller than the equilibrium
value. On the other hand, memory effects of the critical
regime are preserved more efficiently than if the system were in equilibrium throughout.
One observes that when $T$ is below $T_{c}$,  
the non-equilibrium  value of $\xi$ at that temperature is larger than
the equilibrium value. Similar observations were made previously in Ref.~\cite{Berdnikov:1999ph}.

Turning now to the evolution of the non-Gaussian cumulants $S$ and $K$, we first recall that in 
equilibrium $\k^{\equ}_{3}$, or equivalently $S^{\equ}$, is an odd function of the Ising variable $h$. It will 
therefore will flip sign when crossing the cross-over line, as demonstrated by the 
dashed curve in Fig.~\ref{fig:traja}~(c)). 
In contrast, $\k^{\equ}_{4}$ or $K^{\equ}$ is an even function of the Ising variable $h$. 
It is negative at the cross-over temperature and positive away from
it as shown in the corresponding dashed curve in Fig.~\ref{fig:traja}~(d). 
However we demonstrate in Fig.~\ref{fig:traja} that 
the non-equilibrium evolution of skewness and kurtosis do not necessarily follow the evolution of the corresponding
equilibrium cumulants, and can be radically different in both their magnitude and sign. 

These differences occur because, as previously noted, the evolution of higher cumulants is coupled to the lower ones. 
Therefore how $K$ (or $S$) evolve will not only depend on its deviation from the equilibrium value, but also on the
non-equilibrium values of other cumulants. As we shall discuss shortly, these deviations off-equilibrium have significant phenomenological implications for the search for a 
critical point in a beam energy scan. Specifically, in Fig.~\ref{fig:traja}, the dashed vertical lines correspond to freeze-out trajectories I, II, and III (left to right), which provide snapshots of the non-equilibrium cumulants that may be measured in experiments. We shall return to a more detailed discussion of these 
in Sec.~\ref{sec:sqrts}.

\section{Towards modeling the RHIC beam energy scan
\label{sec:BES}
}
The results we presented for the non-Gaussian cumulants off-equilibrium potentially strongly impact the interpretation of the 
results of ongoing and future critical point searches with the beam energy scan (BES) at RHIC.  To further explore these, 
we will solve the evolution equation for fixed-$\mu$ trajectories broadly spanning the critical regime. 
In our simple model, this would be the equivalent of varying $\sqrt{s}$.
We will then be able to compute the non-equilibrium cumulants for a given $\tau_{\rel}/\tau_{I}$ for every point in the critical regime. 
\subsection{ Memory effects and the sign of non-equilibrium skewness and kurtosis}
\begin{figure}
\centering
        \includegraphics[width=0.45\textwidth]{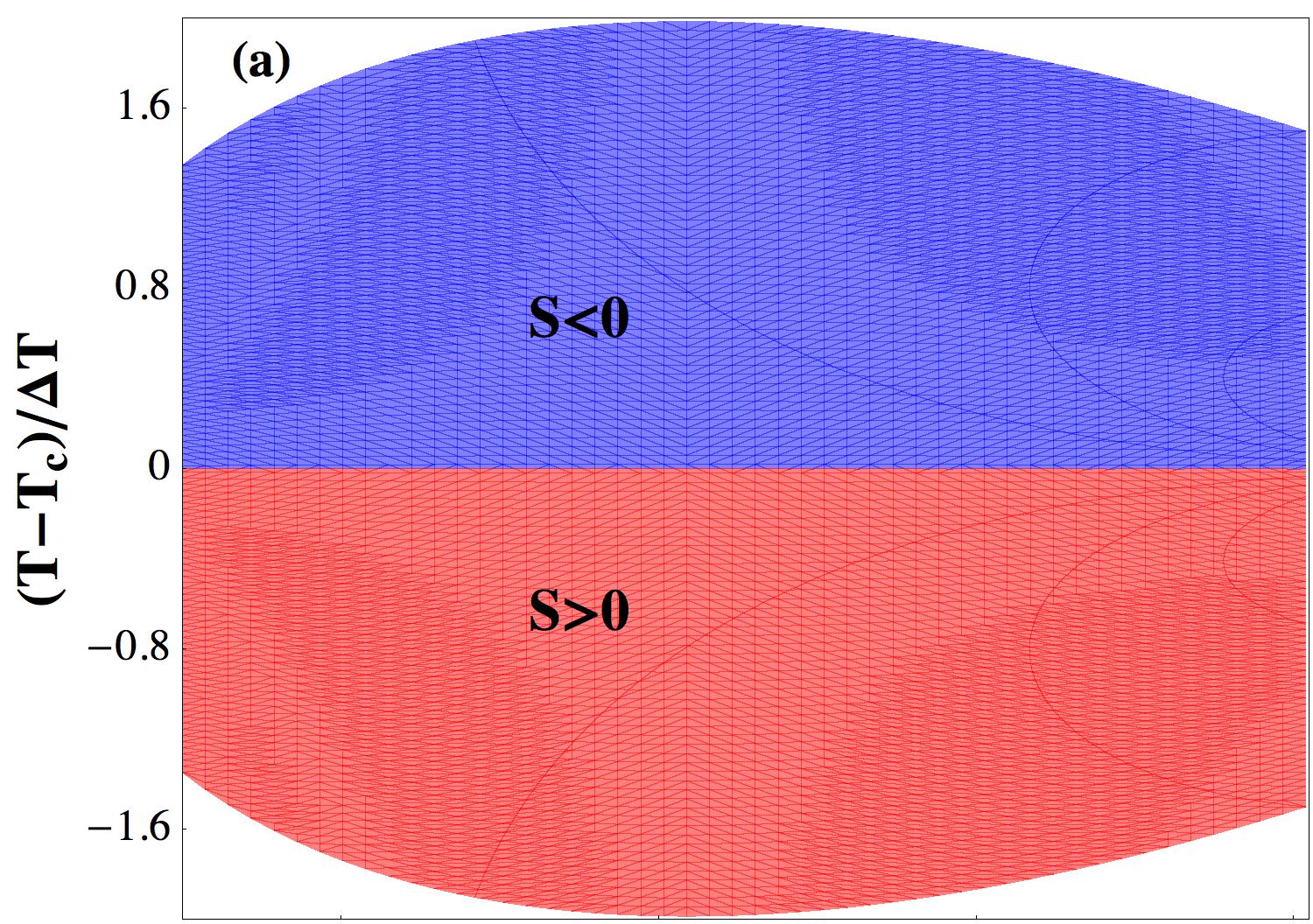}
        \includegraphics[width=0.45\textwidth]{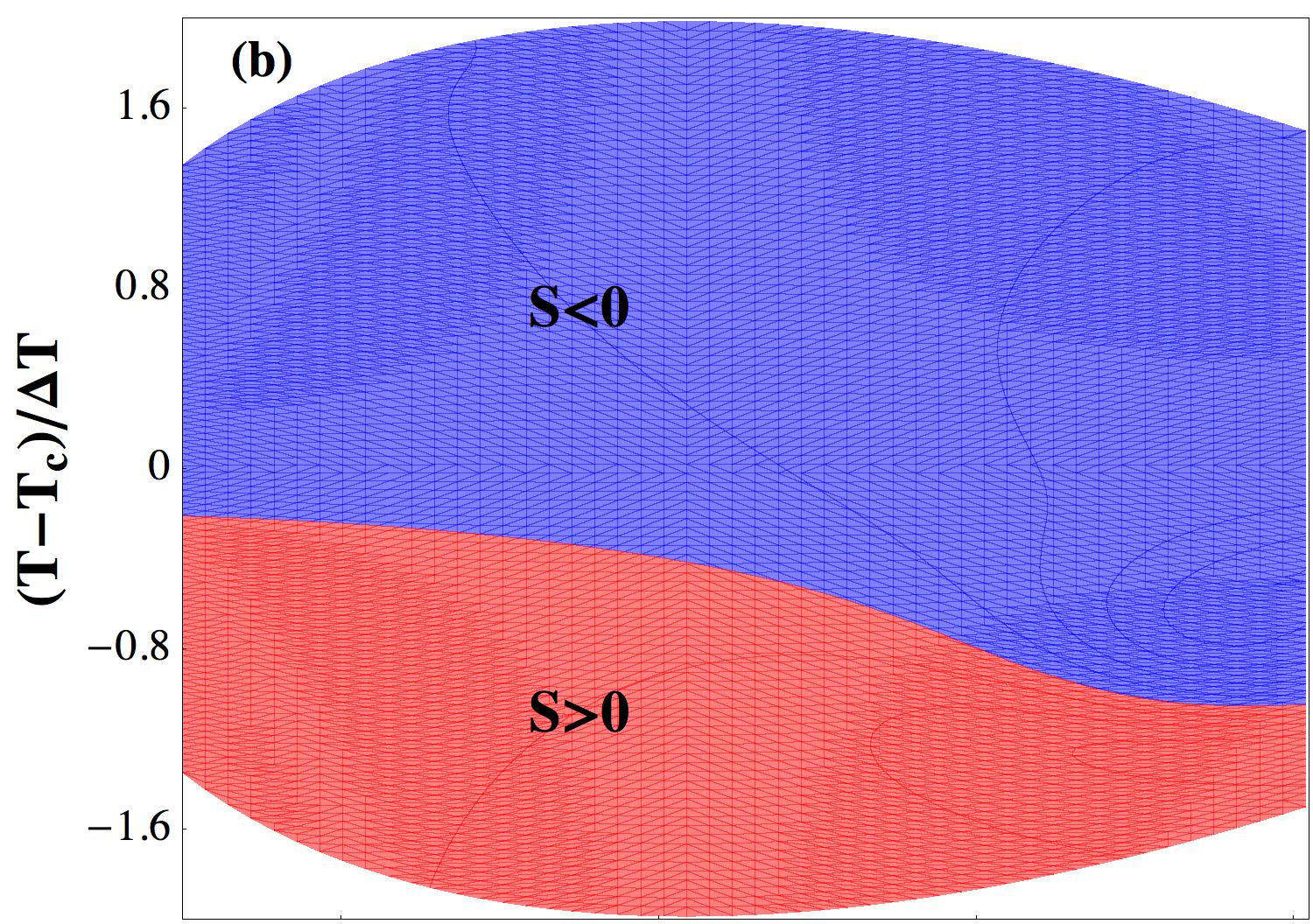}
        \includegraphics[width=0.45\textwidth]{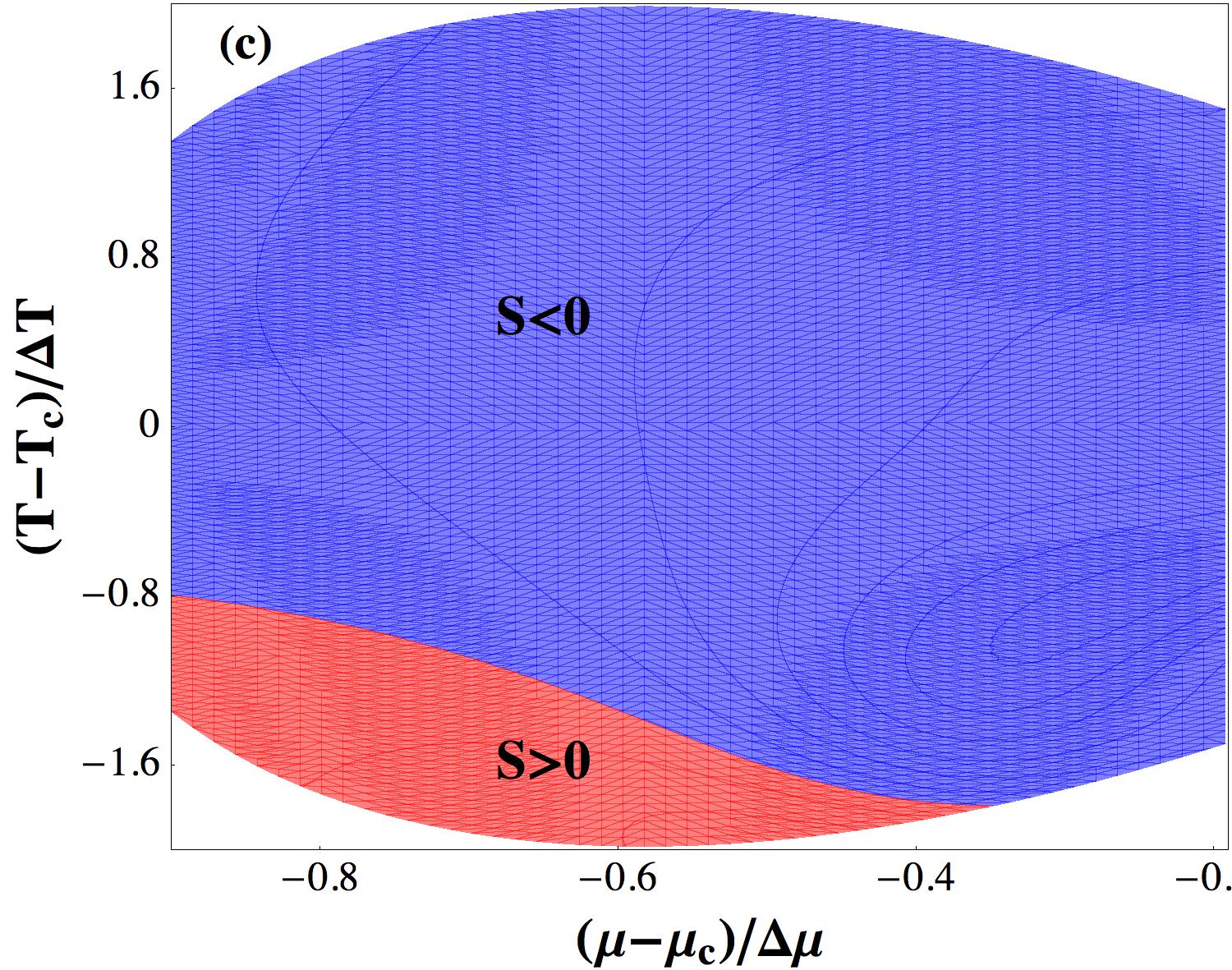}
\caption{
\label{fig:Scontour}(Color online)
Contour plot of equilibrium (top) and non-equilibrium skewness (S) with
$\tau/\tau_{I}=0.05$ (middle) and $\tau/\tau_{I}=0.02$ (bottom).
The $S>0$ region is shown in red and $S<0$ region is shown in blue.
 }
\end{figure}
\begin{figure}
\centering
        \includegraphics[width=0.45\textwidth]{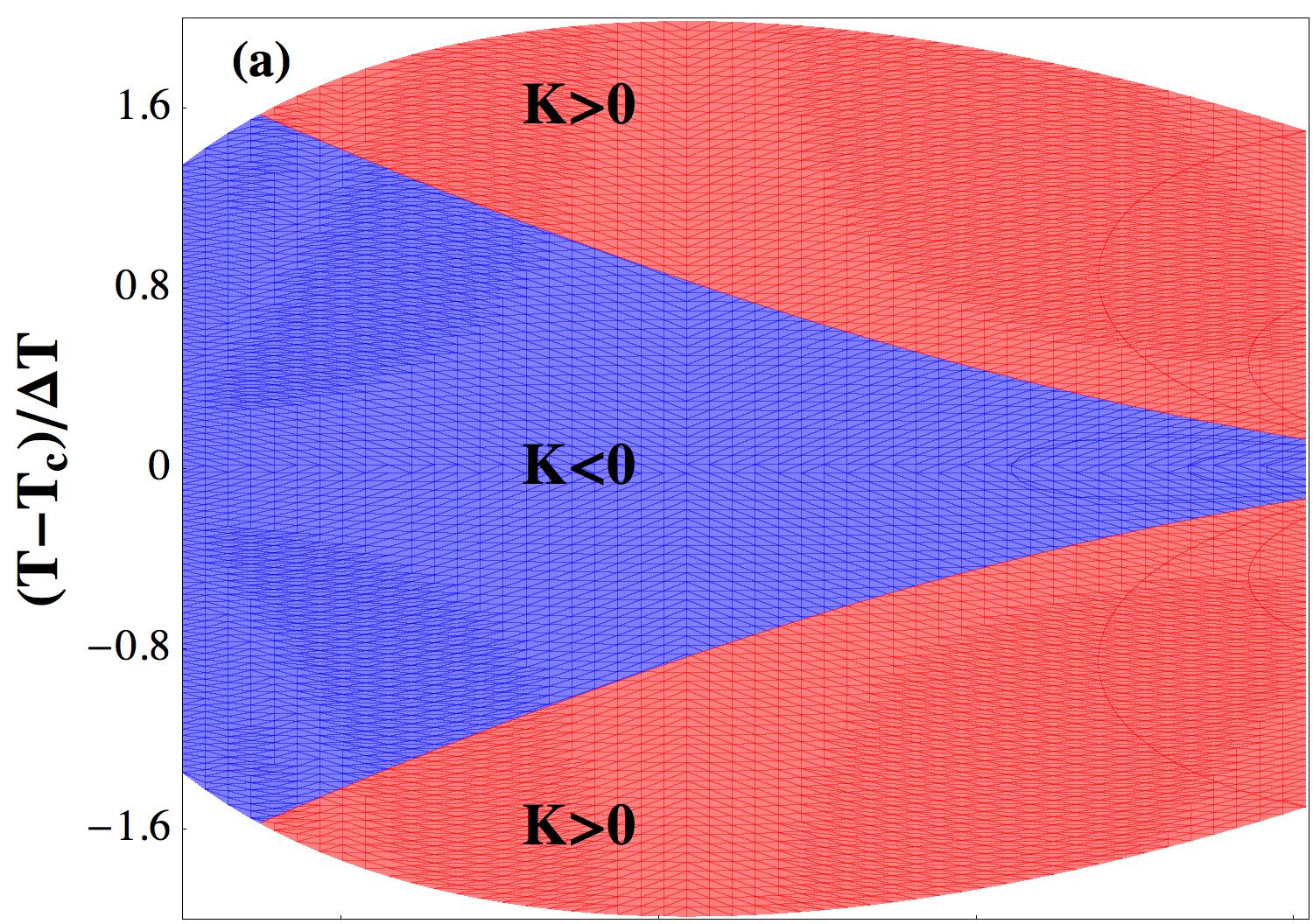}
        \includegraphics[width=0.45\textwidth]{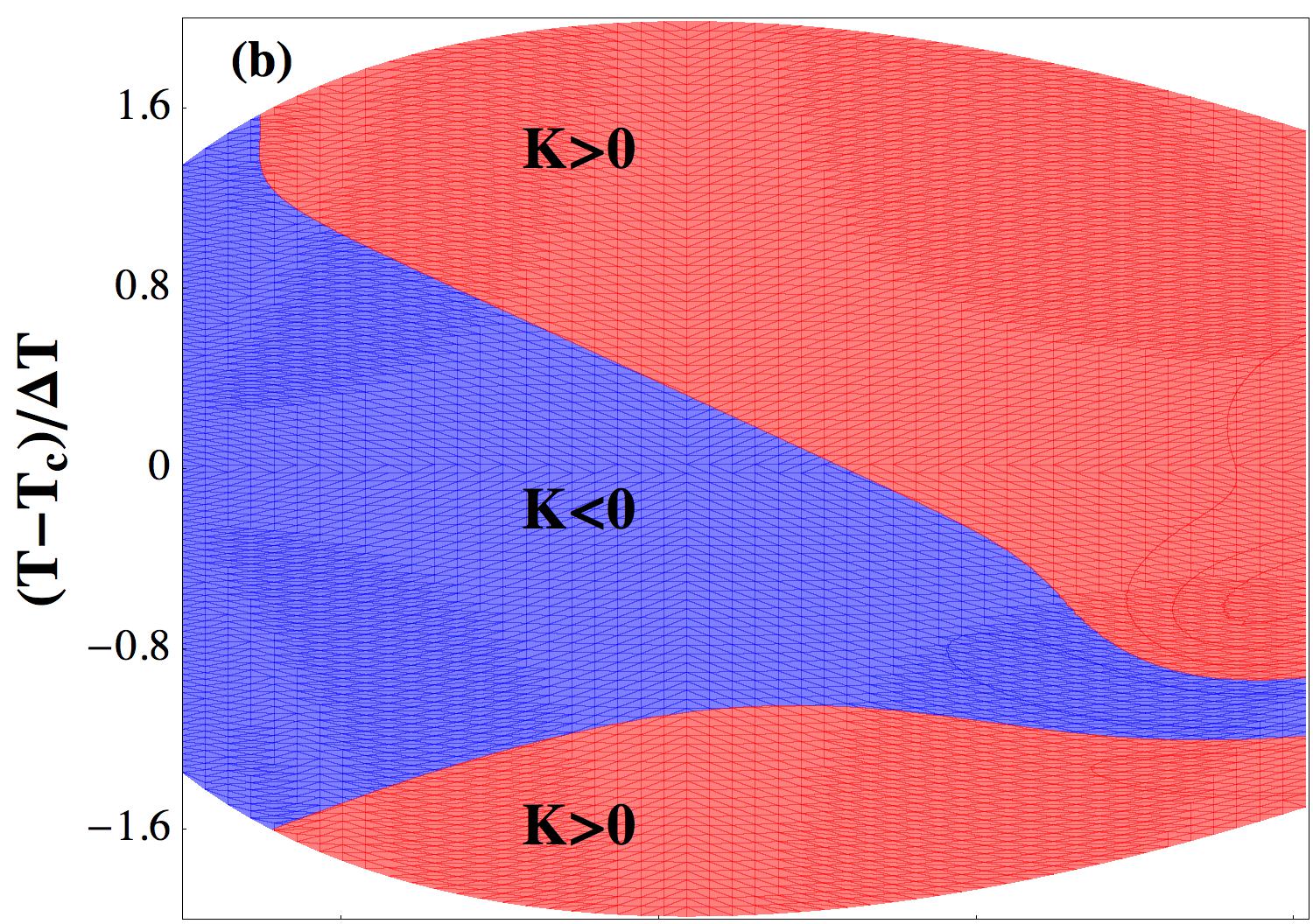}
        \includegraphics[width=0.45\textwidth]{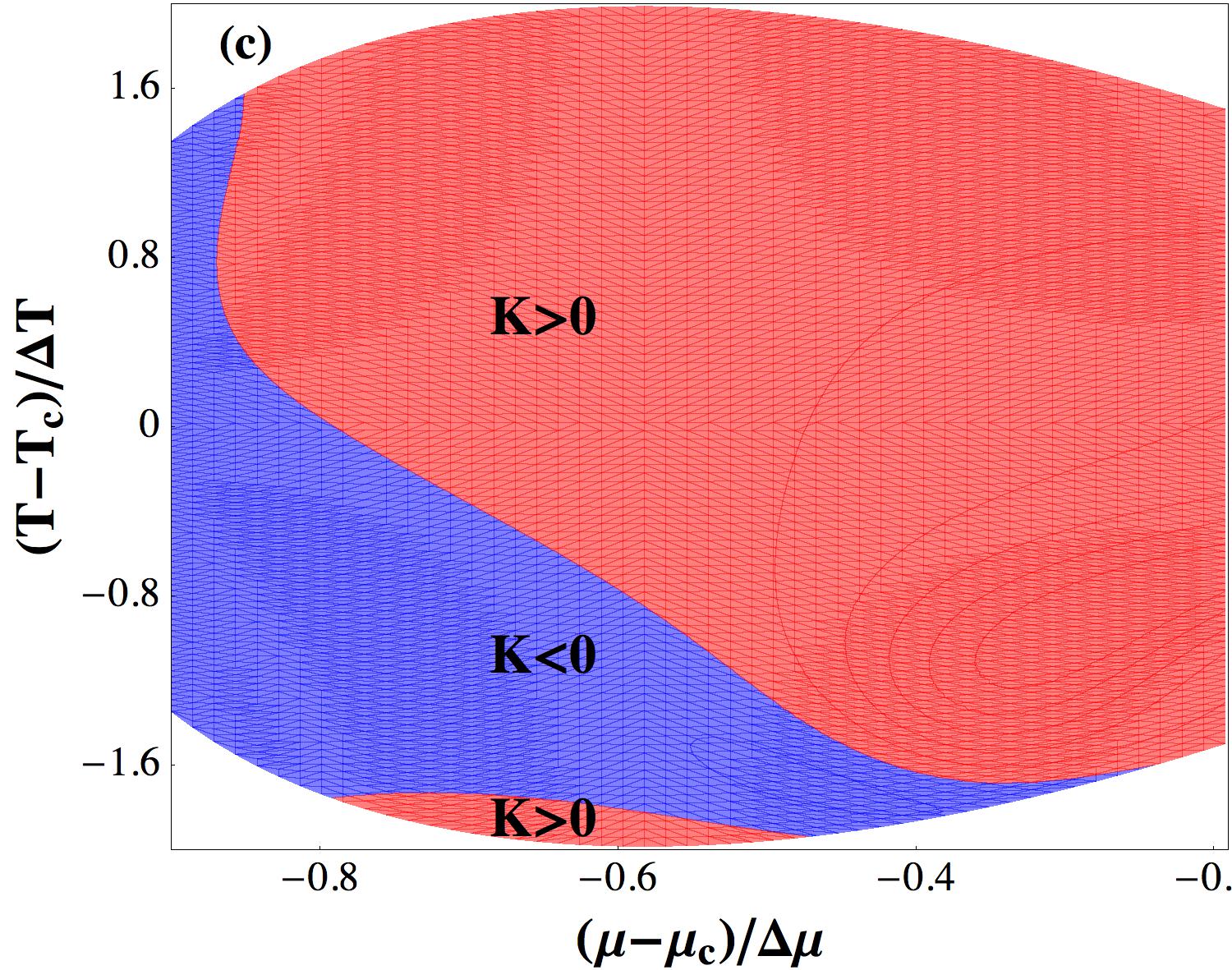}
\caption{
\label{fig:Kcontour} (Color online)
Contour plot of equilibrium (top) and non-equilibrium kurtosis (K) with
$\tau/\tau_{I}=0.05$ (middle) and $\tau/\tau_{I}=0.02$ (bottom).
The $K>0$ region is shown in red and $K<0$ region is shown in blue.
 }
\end{figure}

In Figs.~\ref{fig:Scontour} and \ref{fig:Kcontour} respectively,
we present contour plots for the equilibrium and non-equilibrium
skewness $S$ and kurtosis $K$. Due to memory effects, 
the non-equilibrium contours in the $T-\mu$ plane deform from the corresponding
equilibrium contours; the deformation is enhanced for larger
relaxation times $\tau_{\rm rel}/\tau_{I}$.

We now focus on the sign of  the skewness and
kurtosis, their most prominent feature. 
We illustrate it by plotting regime $S>0$ (or $K>0$) in red and $S<0$ (or $K<0$) in blue.
In equilibrium,
the boundary that separates the regime where $S>0$  and $S<0$ is precisely the cross-over line at $T=T_c$. 
In Fig.~\ref{fig:Scontour}, we fix the sign of the equilibrium skewness in such a way that
$S_{\equ}$ is positive below the cross-over line; this is suggested by the arguments presented in Refs.~\cite{Stephanov:2008qz,Asakawa:2009aj} and the observation 
that the skewness is positive in the hadron resonance gas model. 
 Fig.~\ref{fig:Scontour} demonstrates that for non-equilibrium skewness $S$
the boundary where the skewness changes sign deforms and with increasing $\tau_{\rel}/\tau_{I}$ becomes negative in a larger portion of the area below the cross-over line.  
This is to be expected as the non-equilibrium cumulants carry more
information at early times when the equilibrium skewness is negative; 
a larger relaxation time $\tau_{\rm rel}$ would give more weight to 
early time contributions.
Similarly, as Fig.~\ref{fig:Kcontour} shows, the boundary separating the regime $K>0$ and $K<0$ also deforms. 

With Figs.~\ref{fig:Scontour} and \ref{fig:Kcontour} in mind, 
we may ask how the sign of the non-equilibrium skewness and kurtosis behaves 
as a function of $\mu$ (or $\sqrt{s}$) on the freeze-out curve of a heavy ion collision in the $T$-$\mu$ plane.
As Fig.~\ref{fig:Scontour} illustrates,
the skewness on the freeze-out curves can be either negative
or positive depending on the magnitude of $\tau_{\rel}/\tau_{I}$ and the relative
position of the freeze-out curves. However, despite memory effects,
the sign of \textit{non-equilibrium kurtosis} will still switch from negative to positive when $\mu$ 
approaches $\mu_{c}$ along the direction of cross-over. This change in sign of the $\mu$ dependence of kurtosis is
 insensitive both to the magnitude of $\tau_{\rel}$ as well as the location of the freeze-out curves.   

\subsection{Dependence of non-equilibrium cumulants on $\sqrt{s}$ and freeze-out curves
\label{sec:sqrts}  
  }

We shall now illustrate a few possible experimental outcomes for the behavior of cumulants as a
function of $\mu$ (or $\sqrt{s}$) if the cross-over side of critical regime of QCD
phase diagram is scanned. We emphasized previously that the cumulants of the critical field $\s$ itself are not directly
observable. However the critical field $\s$ is coupled to the net baryon number density and therefore critical fluctuations contribute to the moments of net baryon number
fluctuations that are measured in experiments. Indeed, such contributions are proportional to the corresponding
moments of the $\s$ field
itself~\cite{Stephanov:2008qz,Adamczyk:2014fia}. We may therefore
expect that the $\mu$-dependence of cumulants of the critical fields, as determined in our model, will capture the qualitative behavior of cumulants of particle fluctuations.  

\begin{figure}
        \includegraphics[width=0.45\textwidth]{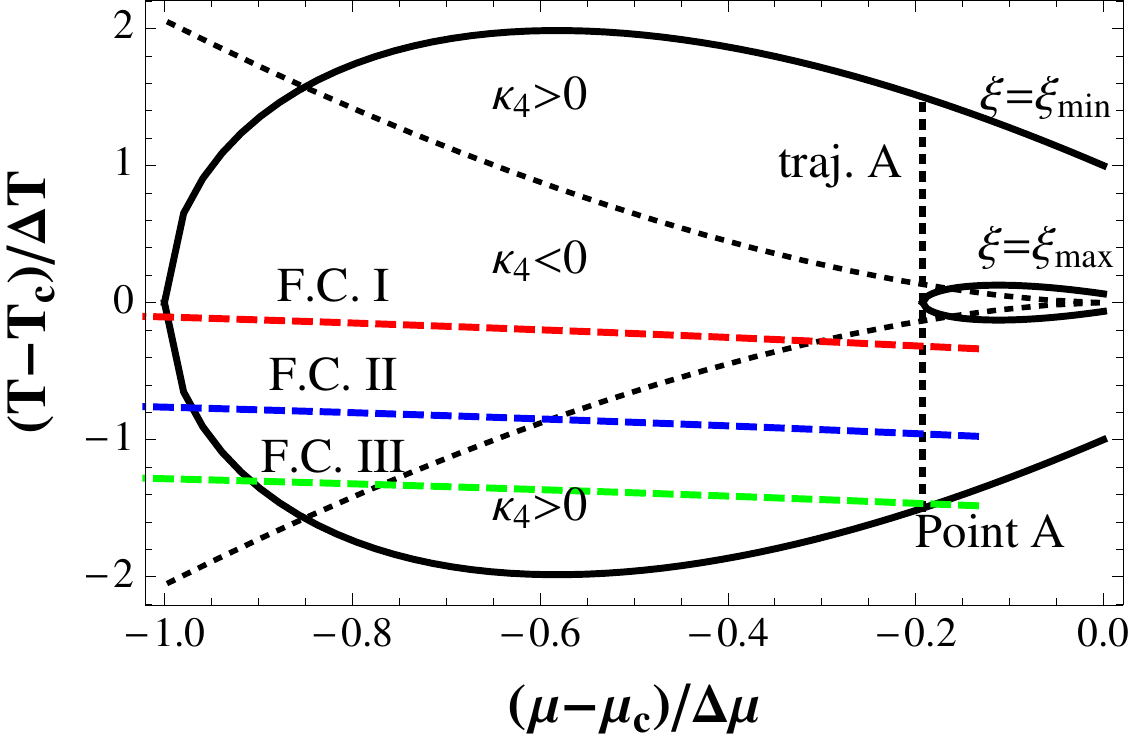}
              \caption{
      \label{fig:BES2}
A sketch of the cross-over side of the scaling regime in the $T, \mu$ plane.
The dotted curves show where the equilibrium value of the kurtosis
$K$ changes sign. The trajectory A  (see text) is shown with the black dashed vertical line. 
Different types of freeze-out curves (F.C.) are shown in red (upper ), blue (middle) and
green (lower) dashed curves, corresponding to type I, II, III respectively (see text).
}
\end{figure}
Consider Fig.~\ref{fig:BES2}, where we have now re-plotted Fig.~\ref{fig:BES} performing the map from $r,h$ variables to $T$ and $\mu$ using \eq\eqref{eq:Tmu_mapping}. 
In this figure, we have superposed the 
trajectories corresponding to three freeze-out curves. Each different choice of a freeze-out curve corresponds to taking a different snapshot of the
evolution of cumulants  as represented by the dashed vertical curves in Fig.~\ref{fig:traja}). Our results will therefore depend on the relative overlap between the QCD
critical regime and the particular freeze-out curve in the QCD phase diagram.
To describe the freeze-out curves,  
we take an empirical parametrization of the heavy-ion collision data from
Ref.~\cite{Cleymans:2005xv},
\be
\label{eq:Tf}
T_f(\mu)
= a-b\,\mu^2_{B} - c\, \mu^4_{B}\, ,
\ee
with $a=0.166~\GeV$, $b=0.139~\GeV^{-1}$, $c=0.053~\GeV^{-3}$.
Given the mapping \eqref{eq:Tmu_mapping}, the overlap between the critical regime and the freeze-out curves depends on
$\mu_{c}, T_{c}$ as well as $\Delta T, \Delta \mu$.
Currently the location of the QCD critical point and the width of QCD critical
regime are not known. Model calculations~\cite{Hatta:2002sj} and lattice QCD calculations~\cite{Gavai:2008zr} suggest
that $\Delta \mu\approx 0.1~\GeV$. 
We therefore set $\mu_c=0.25~\GeV,\Delta \mu= 0.1~\GeV, \Delta T/T_c=1/8$.
Consequently, 
the overlap between a freeze-out curve and QCD critical regime will depend on $T_c$.
In practice, we shall take three different values, $T_{c}=0.165,0.18,0.194~\GeV$, to
represent three freeze-out trajectories overlapping with the critical regime:
I) the freeze-out curve is near the cross-over line, 
II) below the cross-over line and III) near the edge of the critical regime. 
They are plotted in Fig.~\ref{fig:BES2} and will be labeled I, II, III
respectively\footnote{A reader might question whether the values of $T_c$ considered are reasonable given our present knowledge of the QCD phase diagram. However, for 
our illustrative purposes, the choice of $T_c$ is just an easy way to represent the distance of the critical point from the freeze-out trajectory. Alternative ways of representing this relative 
separation are certainly feasible; for instance, we could have chosen smaller values of $T_c$ and modified the parametrization of the freeze-out curves.}
\begin{figure}
\centering
        \includegraphics[width=0.45\textwidth]{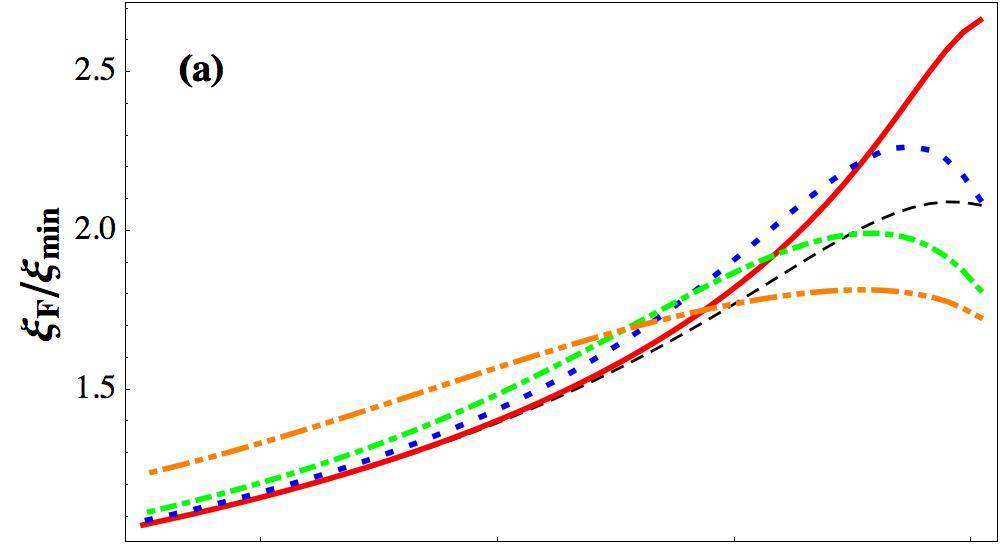}
        \includegraphics[width=0.45\textwidth]{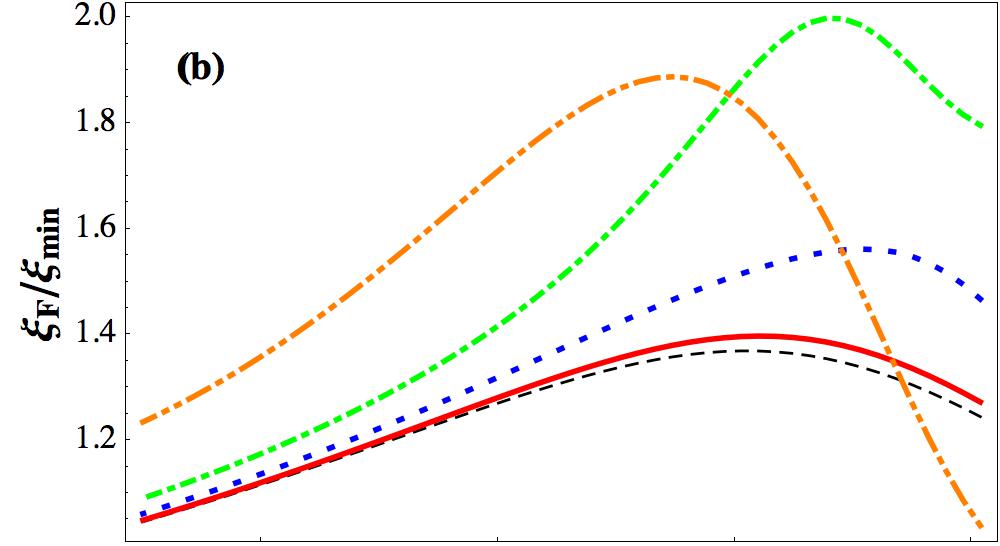}
        \includegraphics[width=0.45\textwidth]{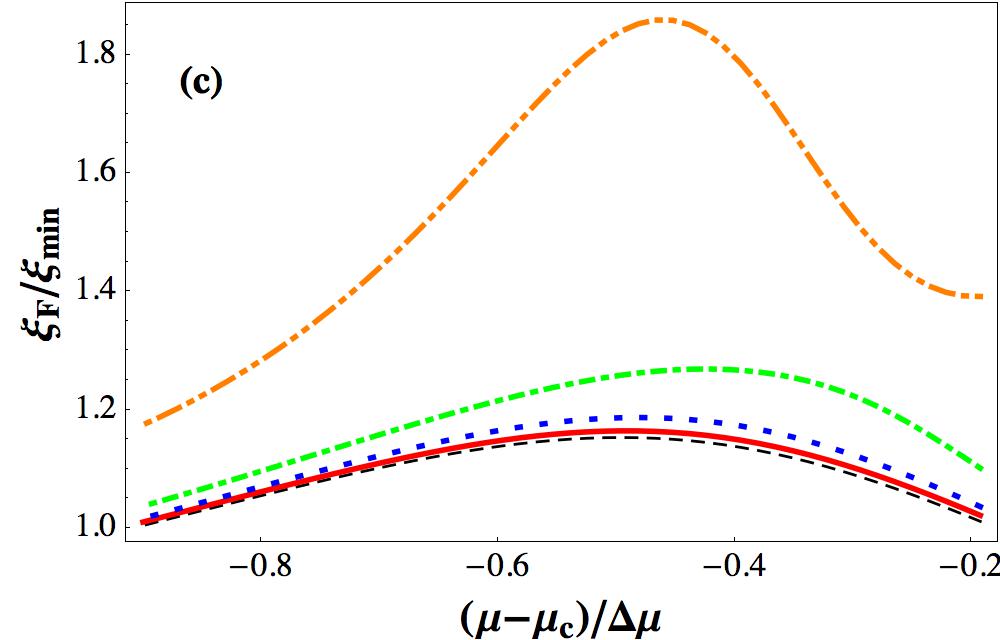}
\caption{
\label{fig:FCV}
(Color online)
The non-equilibrium values of the effective correlation length
$\xi_{F}$ on the freeze-curves
as a function of $\mu$. 
The results corresponding to freeze-out curves of type I, II, III
 (as shown in Fig.~\ref{fig:BES2}) are (a), (b), (c) respectively. 
Non-equilibrium cumulants with
$\tau_{\rel}/\tau_I=0.005,0.02,0.05,0.2$ are represented by 
solid red, dotted blue, single dot dashed green, double dot dashed orange curves respectively. 
The dashed curves plot the corresponding equilibrium values--where not visible, they fully overlap with the solid red curve. 
All results are normalized by the corresponding equilibrium value at
the end point of trajectory A  as shown in Fig.~\ref{fig:BES2}.
}
\end{figure}

\begin{figure}
\centering
        \includegraphics[width=0.45\textwidth]{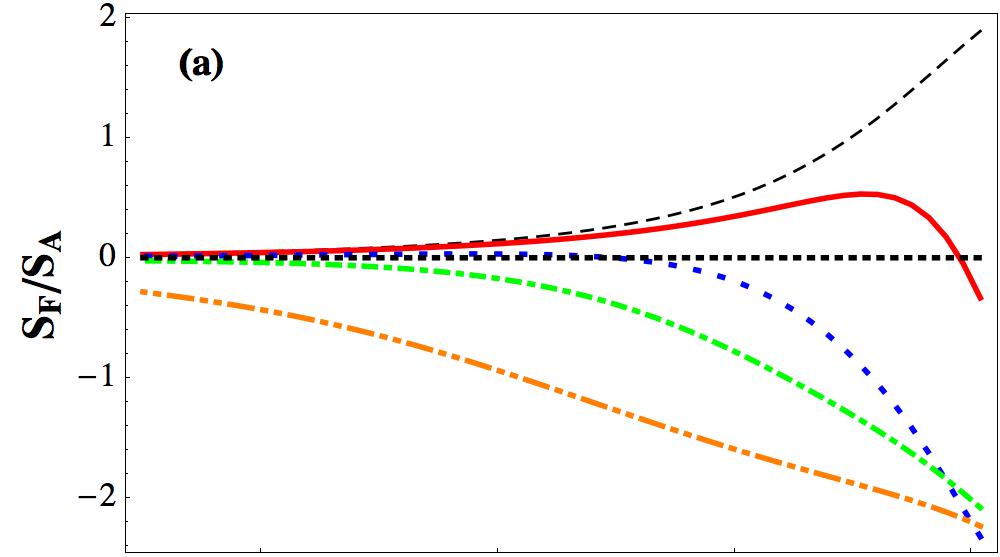}
        \includegraphics[width=0.45\textwidth]{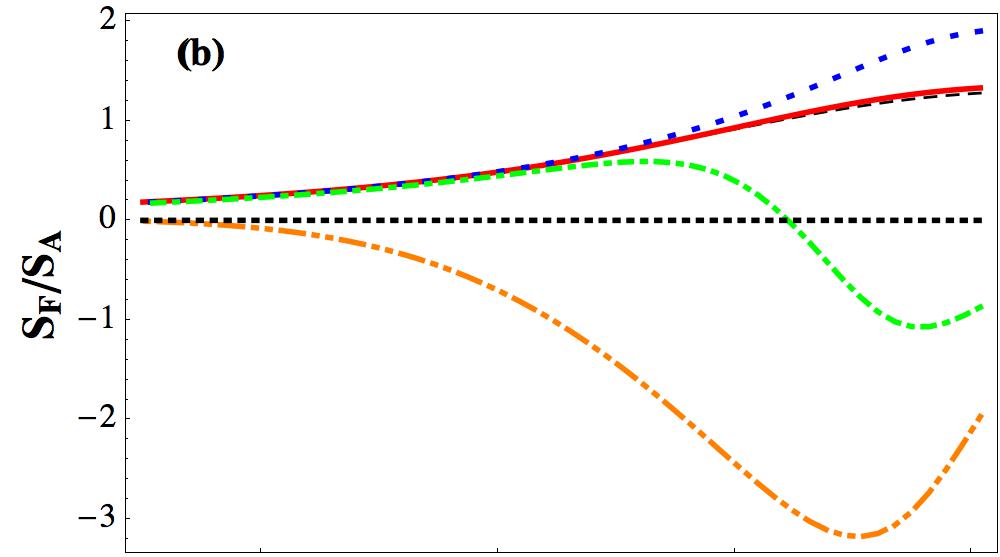}
        \includegraphics[width=0.45\textwidth]{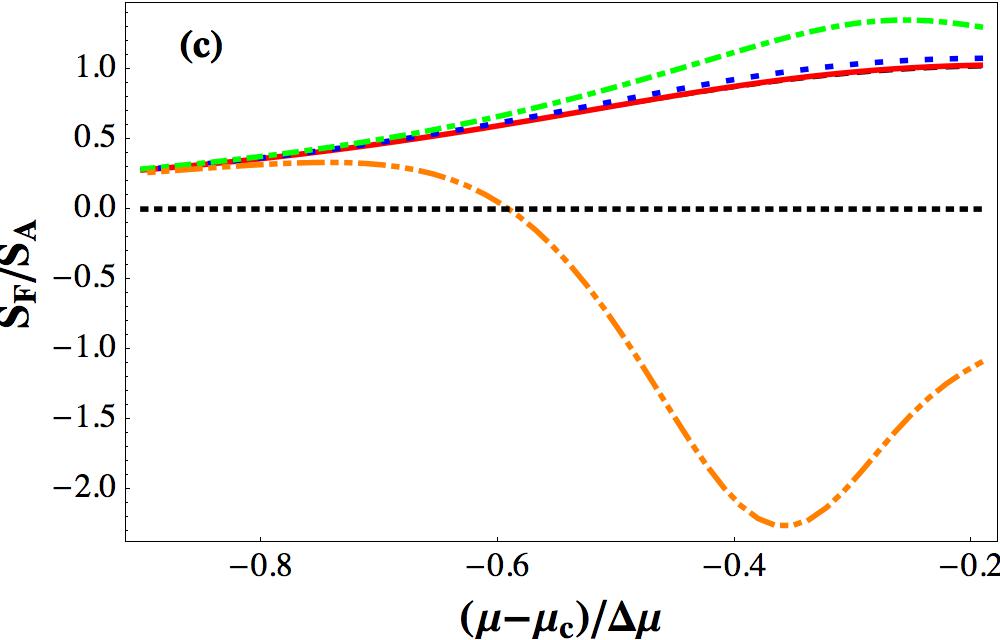}
\caption{
\label{fig:FCS}
(Color online)
The non-equilibrium values of the non-Gaussian cumulant
$S_{F}$ on the freeze-curves
as a function of $\mu$. 
The results corresponding to freeze-out curves of type I, II, III
 (as shown in Fig.~\ref{fig:BES2}) are (a), (b), (c) respectively. 
Non-equilibrium cumulant results for 
$\tau_{\rel}/\tau_{I}=0.005,0.02,0.05,0.2$ are displayed with 
solid red, dotted blue, single dot dashed green, double dot dashed orange curves respectively.  
The dashed curves plot the corresponding equilibrium values. 
All results are normalized by the corresponding equilibrium value at
the end point trajectory A, as shown in Fig.~\ref{fig:BES2}).
}
\end{figure}

\begin{figure}
\centering
        \includegraphics[width=0.5\textwidth]{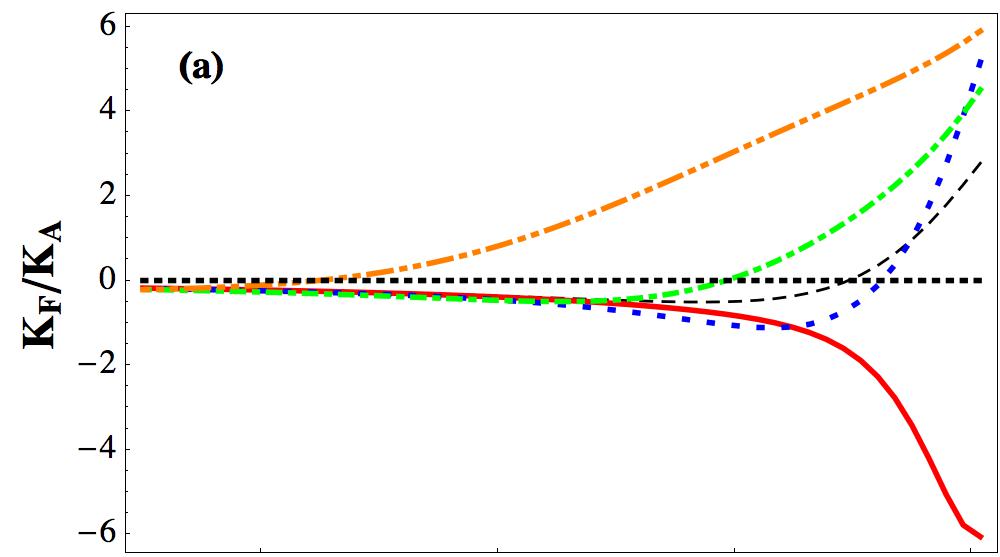}
        \includegraphics[width=0.5\textwidth]{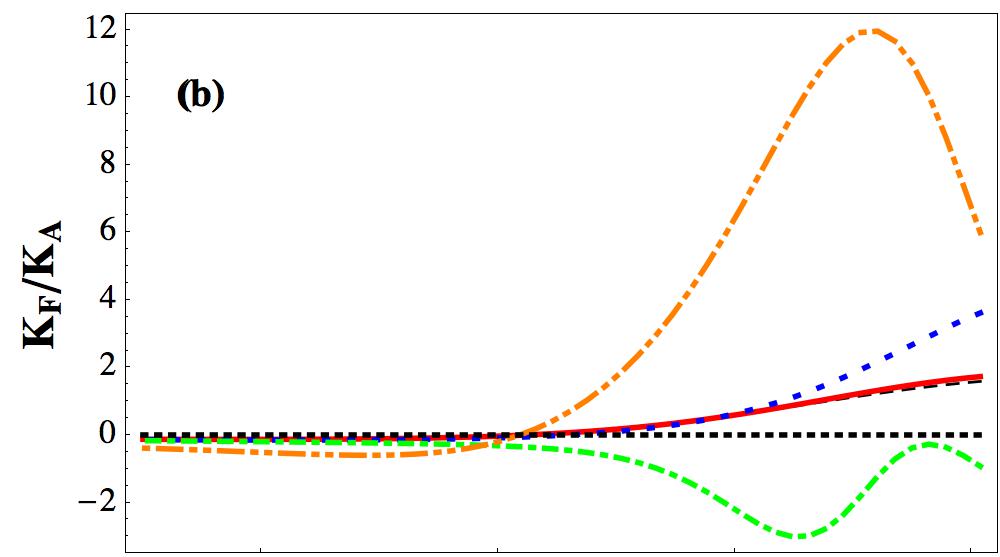}
        \includegraphics[width=0.5\textwidth]{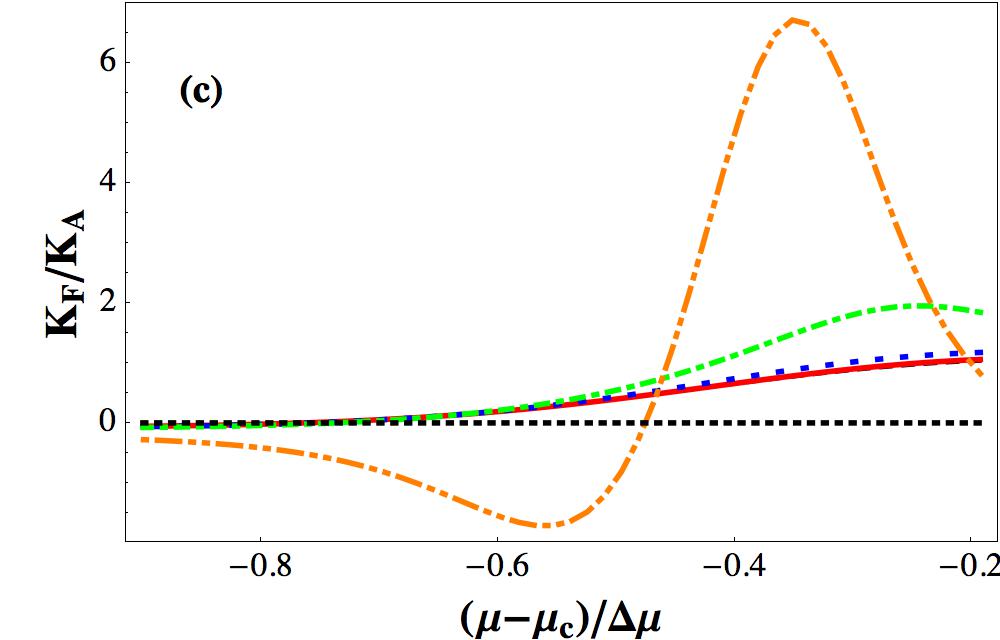}
\caption{
\label{fig:FCK}
(Color online)
The non-equilibrium values of the non-Gaussian cumulant
$K_{F}$ on the freeze-curves as a function of $\mu$. 
The results corresponding to freeze-out curves of type I, II, III
 (as shown in Fig.~\ref{fig:BES2}) are (a), (b), (c) respectively. 
Non-equilibrium cumulant results for 
$\tau_{\rel}=0.005,0.02,0.05,0.2\tau_{I}$ are displayed with 
solid red, dotted blue, single dot dashed green, double dot dashed orange curves respectively. 
The dashes curves plot the corresponding equilibrium values. 
All results are normalized by the corresponding equilibrium value at
the end point trajectory A, as shown in Fig.~\ref{fig:BES2}).
 }
\end{figure}

For each constant $\mu$, we will denote the values of the non-equilibrium cumulants at the point
where the trajectory intersects the freeze-out curves by the subscript ``F''.
We will then study the dependence of the cumulants on freeze-out curves as
a function of $\mu$. Our results for $\xi_{F} (\mu),
S_{F}(\mu),K_{F}(\mu)$ are shown in
Figs.~\ref{fig:FCV},~\ref{fig:FCS},~\ref{fig:FCK} respectively.
We observe generally that the  $\mu$ dependence of the non-equilibrium cumulants can be different from the
equilibrium cumulants (which are represented by dashed lines).
As we anticipated previously, the values of the non-equilibrium $\xi$ on all the freeze-out curves are considerably amplified for a wide range
of $\tau_{\rel}/\tau_{I}$. This implies that even if the freeze-out curve is located at the edge
of the critical regime (as for the freeze-out curve III shown in Fig.~\ref{fig:FCV}~(c)), memory effects ensure that the signature of critical fluctuations is not necessarily suppressed. Regarding skewness, we noted arguments advanced that the equilibrium skewness is positive below the cross-over line. 
Our results in Fig.~\ref{fig:FCS} clear demonstrate that memory effects can modify the sign of skewness off-equilibrium to be opposite to that of 
equilibrium skewness. Similar deviations from equilibrium expectations are observed for kurtosis, as shown in Fig.~\ref{fig:FCK}

\begin{figure}
\centering
            \includegraphics[width=0.45\textwidth]{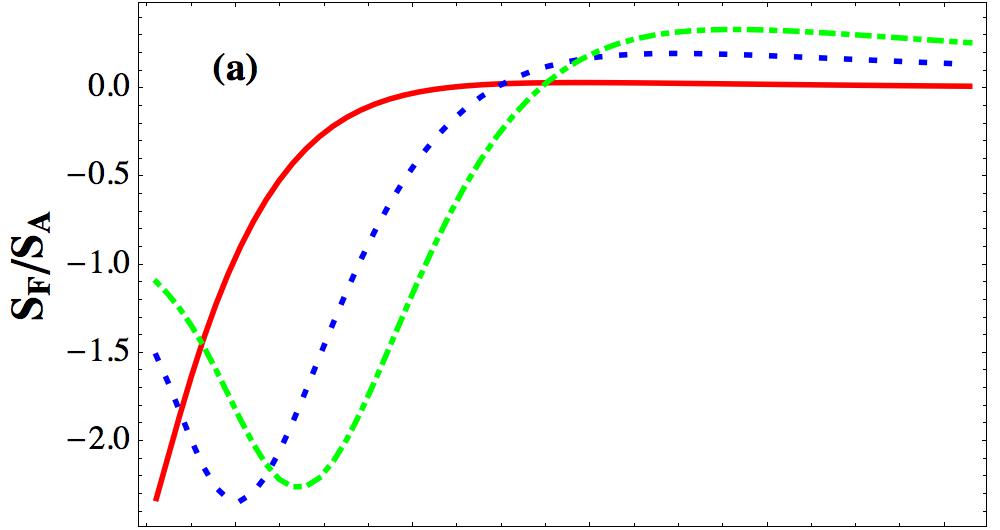}  
        \includegraphics[width=0.45\textwidth]{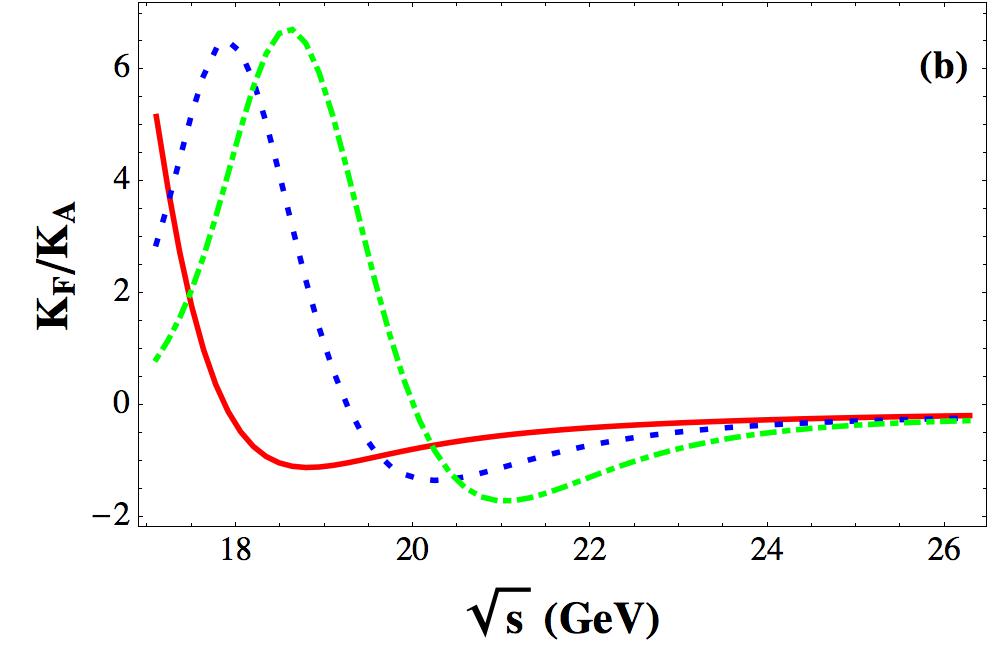}
 \caption{
\label{fig:SKcompare}
 (Color online) The $\sqrt{s}$-dependence (assuming \eq\eqref{eq:sf}) of non-equilibrium non-Gaussian cumulants on the
freeze-curves with different choices of the relative position of freeze-out curves and $\tau_{\rel}$.
(Top)  skewness
$S_{F}$ vs $\sqrt{s}$; (Bottom) kurtosis $K_{F}$ vs $\sqrt{s}$. 
The results obtained from the combination, i.e.,  
(F.C. I, $\tau_{I}/\tau_{\rel}=0.02$), 
(F.C. II, $\tau_{I}/\tau_{\rel}=0.05$), (F.C. III, $\tau_{I}/\tau_{\rel}=0.2$), 
are displayed in solid red, dotted blue, single dot dashed green curves respectively. 
  }
\end{figure}

From Figs.~\ref{fig:FCS} and ~\ref{fig:FCK},
we observe that the non-Gaussian cumulants $S_{F} (\mu)$ and $K_{F} (\mu)$ are sensitive to the relative positions of freeze-out curves (or $T_{c}$) and $\tau_{\rel}$. 
We can go one step further, and given a model for the $\sqrt{s}$
dependence of $\mu$, examine how these fluctuations 
 vary as a function of $\sqrt{s}$. Ref.~\cite{Cleymans:2005xv}, from fits to data, developed the parametrization, 
\be
\label{eq:sf}
\mu(\sqrt{s})= \frac{d_0}{d_1\sqrt{s}+1}\, , 
\ee
where $d_0=1.308~\GeV$  and $d_{1}=0.273~{\rm GeV}^{-1}$.
Using this relation, we obtain the results shown in Fig.~\ref{fig:SKcompare} for the skewness and kurtosis respectively. In particular, we demonstrate that very similar curves, as a function of $\sqrt{s}$ can be obtained by different combinations of freeze-out curves 
and relaxation times. These results suggest that great care should be exercised in interpreting trends as a function of $\sqrt{s}$ of measured non-Gaussian cumulants. While these may signify the onset of critical dynamics, more needs to be done to model freeze-out conditions and constrain $\tau_{\rm rel}$ before definitive statements can be made regarding the discovery of a critical point. 

\section{
Summary and Conclusion
\label{sec:summary}
}
 We derived in this paper a set of equations Eqs.~\eqref{eq:kappa_evo} that describe the evolution of non-equilibrium cumulants of the critical field $\s$ in the QCD critical regime. In particular, we obtained novel expressions for the off-equilibrium evolution of non-Gaussian cumulants.  With equilibrium initial conditions given by the three-dimensional Ising model (which belongs to the static universality class of the QCD critical point), we applied these equations to study
the real time dynamics of cumulants along a trajectory passing through the cross-over side of the critical regime. Since this is a first exploratory study, our studies were performed in 
the framework of a very simple model of expansion dynamics in heavy-ion collisions. With further data motivated parametrizations of chemical freeze-out in the $T-\mu$ plane, we were 
able to  obtain snapshots of the non-equilibrium cumulants along freeze-out curves that overlapped to differing extents with the QCD critical region.

Our chief conclusion is that memory effects are responsible for substantial differences in the temporal evolution of non-equilibrium and equilibrium cumulants. 
Specifically we found that the sign of non-equilibrium skewness in the vicinity of the critical point can be opposite to the value expected in equilibrium. 
Similar conclusions are suggested by our results for non-equilibrium kurtosis. Our results are sensitive to the relaxation time $\tau_{\rm rel}$ of the critical field $\s$ fluctuations and the 
location of the freeze-out curves in the QCD phase diagram relative to the extent of the QCD critical region. We find that memory effects are important to such an extent that even trajectories that traverse the edge of the critical regime are sensitive to its dynamics. 

Our treatment of critical fluctuations in the QCD critical regime was strongly motivated by the description of similar fluctuations in the three-dimensional Ising model. In the latter case, cumulants of critical fluctuations are expressed in terms of the reduced temperature $r$ and the rescaled magnetic field $h$. The map of the description of critical fluctuations in terms of $r$ and $h$ to $T$ and $\mu$ is non-universal, and is a significant source of systematic uncertainty in treatments of critical dynamics in the QCD critical regime. This  uncertainty, coupled with our ignorance of $\tau_{\rm rel}$, provide fundamental obstacles to quantitative studies of real time critical dynamics in QCD. 

Indeed, because of the importance of  non-equilibrium effects, lattice
studies of equilibrium cumulants, while of fundamental importance, 
may not be sufficient. These must be accompanied by progress in non-equilibrium studies of the QCD critical regime. One promising approach is the use of classical statistical real time simulations~\cite{Berges:2012ty,Mesterhazy:2015uja} that have also previously been applied to studying the non-equilibrium dynamics of the very earliest stages of high energy heavy-ion collisions~\cite{Berges:2013eia,Epelbaum:2013waa}. Detailed dynamical models of the space-time evolution of heavy-ion collisions as a function of beam energy are also very important. In particular, models that build in the transport of conserved charges and reproduce bulk features of these collisions such as particle spectra can place strong constraints on the parameter space for the non-equilibrium evolution of cumulants. 

In this work, we have concentrated on critical dynamics on the cross-over side of the critical regime. From the perspective of a critical point search, this approach is appropriate because it is easier in both experiments and in lattice gauge theory computations to extend explorations of the QCD phase diagram starting from the regime of high temperatures and low baryon chemical potentials. However, if a critical point is localized, it would be of great interest to understand non-equilibrium dynamics on the first-order side of phase diagram. In this regard, applying the framework discussed here from the cross-over critical regime to the first-order critical regime of the QCD phase diagram is a useful extension to be pursued in future studies. 

\acknowledgments

We would like to thank Frithjof Karsch, Krzysztof Redlich, Bjoern Schenke and Misha Stephanov for very valuable discussions
and Krishna Rajagopal for detailed and constructive comments on the manuscript. This work was supported by 
DOE Contract No. DE-SC0012704.

\begin{appendix}

\section{Parametric representation of equilibrium cumulants in
 the Ising critical regime
\label{sec:EOS}  
  }
In this section,
we explain the parameterization of the equilibrium cumulants
$M^{\equ}(r,h),\k^{\equ}_{n}(r,h), n=2,3,4,\ldots$ in the critical regime in terms of the Ising
variables $r$ and $h$ used in this paper.
For this purpose,
we only need to know the equilibrium magnetization $M^{\equ}(r,h)$ as
equilibrium cumulants can be computed by taking derivatives of
$M^{\equ}(r,h)$ with respect to $h$ at fixed $r$,
\be
\label{eq:M_kappa}
\k^{\equ}_{n+1} =\frac{1}{(V_4 H_{0})^{n}}
\(\frac{\pd^{n}M^{\equ}(r,h)}{\pd h^{n}}\)_{r}\, .
\qquad
n=1,2,3,\ldots
\ee
Here $H_{0}$ is a dimensionful parameter (of mass dimension $3$) which relates reduced
magnetic field $h$ to the un-reduced magnetic field.

To parametrize $M^{\equ}(r,h)$,
we use  the linear parametric model \cite{PhysRevLett.23.1098,ZinnJustin:1999bf}.
In this parametrization,
one introduces two new variables $R,\theta$ which are related to (dimensionless) Ising
variable $r, h$ as
\be
\label{eq:rh_Rtheta}
r(R,\theta)= R(1-\theta ^2)\, ,
\qquad
h(R,\theta)= \Delta h\, R^{\b \d}\,\th(\theta)\, ,
\ee
Following Ref.~\cite{Stephanov:2011pb},
we will use
\be
\label{eq:th}
\th(\theta) = 3\theta \[1-\(\frac{(\delta-1)(1-2\beta)}{(\delta-3) }\)\theta^2\]\, .
\ee
Here $\b,\d$ are standard critical exponents and we will use the values
obtained from mean field theory, $\b=1/3,\delta = 5$. 
In these $R,\theta$ variables, $\theta=0$ corresponds to the crossover
line and $|\theta|=\sqrt{3/2}$ 
corresponds to the coexistence (first order transition) line. The  equilibrium ``magnetization'' $M^{\equ}_{0}(r,h)$(or $\s_{0}$)
is given by
\be
\label{eq:M_Rtheta}
M^{\equ}(R,\theta)
= M_{0}R^{\b}\theta\, ,
\ee
where $M_{0}$ sets the scale of ``magnetization''. 
The parametrization introduced describes the equation of state with a precision
sufficient for our purpose.

We now compute $\k^{\equ}_{n}$ using \eq\eqref{eq:M_kappa} and \eq\eqref{eq:M_Rtheta}.
Explicitly, we have
\begin{eqnarray}
\label{eq:kappa_rh}
\k^{\equ}_{2}(R,\theta)
&=&\frac{M_{0}}{V_{4}H_{0}}\frac{1}{R^{4/3}(3+2\theta^2)}\, , \\
\k^{\equ}_{3}(R,\theta)
&=&\frac{-M_{0}}{(V_{4}H_{0})^2}\frac{4\theta(9+\theta^2)}{R^{3}(3-\theta^2)(3+2\theta^2)^3}\, , \\
\k^{\equ}_{4}(R,\theta)
&=&\frac{-12M_{0}}{(V_{4}H_{0})^3}\nonumber \\
&\times &\frac{\(81-783\theta^2+105\theta^4-5\theta^6+2\theta^8\)}
{R^{14/3}(3-\theta^2)^3(3+2\theta^2)^5}\, .
\end{eqnarray}
Finally, we convert $\k^{\equ}_{n}(R,\theta)$ into
$\k^{\equ}_{n}(r,h)$ using \eq\eqref{eq:rh_Rtheta}.
We note that 
$M/M_{A},\x/\x_{\rm min},S/S_{A},K/K_{A}$ as presented in this paper
does not depend on the choice of dimensionful normalization $M_{0},H_{0}$.

\section{Detailed derivation of  Eqs.~\eqref{eq:kappa_evo}
\label{sec:derivation_A}
}

We present here a detailed derivation of Eqs.~\eqref{eq:kappa_evo}. 
It is convenient to introduce the generating function of cumulants,
\be
G(\l;\tau)= \log\[Z(\l;\tau)\]\, .
\qquad
Z(\l;\tau)\equiv \<e^{\l \delta\s}\>\, . 
\ee
where the average $\<\ldots\>$ and $\delta\s$ have been defined in
\eq\eqref{eq:average_define} and \eq\eqref{eq:cumulants_def} respectively.
The cumulants are given by
\be
\k_{n}
= \frac{\pd^{n}G(\l)}{\pd \l^{n}}\bigg |_{\l=0}\, , 
\qquad
n=2,3,4,  \ldots\, .
\ee
We also have $\pd_{\l}G(\l;\tau)|_{\l=0}=0$ as $\<\delta\s\>=0$.
Equivalently,
one could read $\k_{n}$ by Taylor expanding $G(\l;\tau)$ around $\l=0$,
\be
\label{eq:G_Taylor}
G(\l;\tau)= \sum^{\infty}_{n=2}\frac{\l^{n}}{n!}\k_{n}(\tau)\, .
\ee

We now take the derivative of $G(\l;\tau)$ with respect to $\tau$,
\be
\label{eq:Gdtau}
\pd_{\tau}G(\l;\tau)
= \frac{\pd_{\tau}Z(\l;\tau)}{Z(\l;\tau)}
= \sum^{\infty}_{n=2}\frac{\l^{n}}{n!}\pd_{\tau}\k_{n}(\tau)\, . 
\ee
Therefore the evolution equations for $\k_{n}$ can be determined by
evaluating $\pd_{\tau}G(\l;\tau)$.
We consider
\begin{multline}
\label{eq:Zdtau}
\pd_{\tau}Z(\l;\tau)
=\pd_{\tau}\[e^{-\l M}\<e^{\l \s}\>\]\\
=  -\l Z(\l;\tau)(\pd_{\tau}M) +e^{-\l M}\pd_{\tau}\<e^{\l \s}\>\, .
\end{multline}
From \eq\eqref{eq:g_evo},
we have
\be
\label{eq:Zdtau1}
e^{-\l M}\pd_{\tau}\<e^{\l\s}\>
=-\frac{1}{m^2_{\s}\tau_{\eff}}
\[ \l \<e^{\l\delta\s} \O'_{0}(\s)\>- \frac{\l^2}{V_{4}}Z(\l;\tau)\]\, . 
\ee
We can now substitute \eq\eqref{eq:Zdtau1} into \eq\eqref{eq:Zdtau} and then plug the results into
\eq\eqref{eq:Gdtau} to obtain,
\begin{eqnarray}
\label{eq:Gdtau1}
\pd_{\tau}G(\l;\tau)
&=& -\l \pd_{\tau}M-\frac{\l b}{\e \tau_{\eff}}
\Bigg\{ F_{1}(M)
\nonumber\\
&+&
\[\frac{\e\<\delta\s e^{\l\delta\s}\>}{b Z }\]F_{2}(M)+\[\frac{\e^2\<\delta\s^2 e^{\l\delta\s}\>}{b^{2}Z}
\]F_{3}(M)
\nonumber\\
&+&\[\frac{\e^3\<\delta\s^3 e^{\l\delta\s}\>}{b^3Z}F_{4}\]-\l b 
\Bigg\}\, .
\end{eqnarray}
Now the evolutions equations for $M$ and cumulants
$\kappa_{n},n=2,3,4,\ldots$
can be determined by Taylor expanding both sides of the above equation in
powers of $\l$
and comparing coefficients in front of  $\l^{n}$.
At order $\l^n, n=1,2,3,4$,
we have, respectively,
\bes
\label{eq:kappa_evo_Full}
\begin{eqnarray}
\label{eq:kappa1}
 \pd_{\tau} M(\tau)
&=&-\tau^{-1}_{\eff}\(\frac{b}{\epsilon}\)
 \Bigg\{ F_{1}(M)+\frac{\e^2}{2}\(\frac{\k_{2}}{b^2}\) F_{3}(M)\nonumber \\
& +&\frac{\e^4}{6}\(\frac{\k_{3}}{\e b^{2}}\) F_{4}(M)\Bigg\}\, ,
\end{eqnarray}
\begin{eqnarray}
\label{eq:kappa2}
\pd_{\tau}\k_{2}(\tau)
&=&-2\tau^{-1}_{\eff}  (b^2)
\Bigg\{
\[\(\frac{\k_{2}}{b^2}\) F_2(M)-1\]\nonumber\\
&+&\frac{\e^2}{2}\[ \(\frac{\k_{3}}{\e b^{2}}\)  F_{3}(M)+\(\frac{\k_{2}}{b^2}\)^2F_4\]
\nonumber\\
&+&\frac{\e^4}{6}\(\frac{\k_{4}}{\e^{2}b^{4}}\) F_{4}
\Bigg\}\, ,
\end{eqnarray}
\begin{eqnarray}
\label{eq:kappa3}
\pd_{\tau}\k_{3}(\tau)
&=&-3 \tau^{-1}_{\eff}  (\e b^3) \Bigg\{\[\(\frac{\k_{3}}{\e b^{2}}\)  F_2(M)+\(\frac{k_{2}}{b^2}\)^2 F_3(M)\]
\nonumber\\
&+&\frac{\e^{2}}{2}\[\(\frac{\k_{4}}{\e^{2}b^{4}}\) F_{3}(M)+3
    \(\frac{\k_{2}}{b^2}\)\(\frac{\k_{3}}{\e b^{2}}\) F_{4}\]
\nonumber \\
&+&\frac{\e^{4}}{6}\(\frac{\k_{5}}{\e^{3}b^5}\) F_{4}
\Bigg\}\, , 
\end{eqnarray}
\begin{eqnarray}
\label{eq:kappa4}
\pd_{\tau}\k_{4}(\tau)
&=&-4\tau^{-1}_{\eff}
\Bigg\{ \left[ (\frac{\k_{4}}{\e^{2}b^{4}}) F_2(M)+3
(\frac{\k_{2}}{b^2})(\frac{\k_{3}}{\e b^{2}}) F_3(M) \right. 
\nonumber\\
&+&
\left. (\frac{k_{2}}{b^2})^3 F_{4}\right]
+\frac{\e^{2}}{2}
\left[ (\frac{\k_{5}}{\e^{3}b^5}) F_{3}(M) \nonumber \right. \\
&+&
\left.\(3(\frac{k_{3}}{\e
    b^3})^3+4(\frac{\k_{2}}{b^2})(\frac{\k_{4}}{\e^{2}b^{4}})\)
    F_{4}\right]
\nonumber\\
&+&
+\frac{\e^{4}}{6}\(\frac{\k_{6}}{\e^{4}b^{6}}\) F_{4}
\Bigg\}\, . 
\end{eqnarray}
\ees
In deriving these equations, we also used the relations 
\begin{eqnarray}
\frac{\<\delta\s e^{\l\delta\s}\>}{Z}
&=&\pd_{\l}G\, , 
\qquad
\frac{\<\delta\s^2 e^{\l\delta\s}\>}{Z}
=\[\pd^2_{\l}G+(\pd_{\l}G)^2\]\, ,
\nonumber \\
\frac{\<\delta\s^3 e^{\l\delta\s}\>}{Z}
&=&\[\pd^{3}_{\l}G+3(\pd^{2}_{\l}G)(\pd_{\l}G)+(\pd_{\l}G)^3\]\, ,
\end{eqnarray}
and \eq\eqref{eq:G_Taylor}.

Keeping contributions to leading order in $\e$ in \eq\eqref{eq:kappa_evo_Full},
we arrive at Eqs.~\eqref{eq:kappa_evo}. 

We note that Eqs.~\eqref{eq:kappa_evo_Full} would still be closed if we
further include $\e^2$ terms in the evolution equations for $M,
\k_{2},\k_{3}$ (but neglect $\e^2$ terms in \eq\eqref{eq:kappa4}). 
The resulting equations,
which we shall refer as the next to leading order (NLO) evolution equations,
 take full sub-leading contributions in $\e^2$
to the evolutions of $M, \k_{2},\k_{3}$ and part of sub-leading
contributions in $\e^2$ to the evolution of $\k_{4}$ into account,
neglecting the term proportional to $\k_{5}$.
We have checked numerically that the differences in computing 
the non-equilibrium cumulants from solutions of the
Fokker-Planck master equation \eq\eqref{eq:FP1}, relative to
i) solutions of the leading order (LO) evolution
equations  Eqs.~\eqref{eq:kappa_evo} and ii) from NLO evolution equations become smaller and smaller with decreasing
$\e^2$. 
For fixed $\e$, 
the difference is relatively larger for non-Gaussian cumulants. 
In particular,
since the sign of non-Gaussian cumulants $\k_{3},\k_{4}$ is
in-definite, 
we observed numerically that occasionally, 
$\k_{3},\k_{4}$ as determined from Eqs.~\eqref{eq:kappa_evo} would
oscillate around zero for a short period of $\tau$.
Indeed,
such behavior will disappear if one solves NLO evolution equations,
which stabilizes the solutions against these oscillations. 
For this reason, 
the results presented in this paper are determined in practice by solving NLO
evolution equations.


%
\end{appendix}

\bibliography{cumulants}

\end{document}